\documentclass[aps,prl,twocolumn,superscriptaddress]{revtex4-1}

\usepackage{graphicx}
\usepackage{siunitx}
\usepackage{fixmath}
\usepackage{amsmath,amsfonts,amssymb}
\usepackage[colorlinks=true,linkcolor=blue,citecolor=blue]{hyperref}

\begin{document}

\title{Co-existence of classical snake states and Aharonov-Bohm oscillations along graphene \textit{p-n} junctions}


\author{P\'{e}ter Makk}
\thanks{These authors contributed equally}
\email{Peter.makk@unibas.ch}
\affiliation{Department of Physics, University of Basel, Klingelbergstrasse 82, CH-4056 Basel, Switzerland}
\affiliation{Department of Physics, Budapest University of Technology and Economics, Budafoki ut 8, 1111 Budapest, Hungary}

\author{Clevin Handschin}
\thanks{These authors contributed equally}
\affiliation{Department of Physics, University of Basel, Klingelbergstrasse 82, CH-4056 Basel, Switzerland}

\author{Endre T\'ov\'ari}
\affiliation{Department of Physics, Budapest University of Technology and Economics, Budafoki ut 8, 1111 Budapest, Hungary}

\author{Kenji Watanabe}
\affiliation{National Institute for Material Science, 1-1 Namiki, Tsukuba, 305-0044, Japan\\}

\author{Takashi Taniguchi}
\affiliation{National Institute for Material Science, 1-1 Namiki, Tsukuba, 305-0044, Japan\\}

\author{Klaus Richter}
\affiliation{Institut für Theoretische Physik, Universit\"{a}t Regensburg, D-93040 Regensburg, Germany}

\author{Ming-Hao Liu}
\affiliation{Department of Physics, National Cheng Kung University, Tainan 70101, Taiwan}

\author{Christian Sch\"{o}nenberger}
\email{Christian.Schoenenberger@unibas.ch}
\affiliation{Department of Physics, University of Basel, Klingelbergstrasse 82, CH-4056 Basel, Switzerland}

\date{\today}

\begin{abstract}
Snake states and Aharonov-Bohm interferences are examples of magnetoconductance oscillations that can be observed in a graphene \textit{p-n} junction. Even though they have already been reported in suspended and encapsulated devices including different geometries, a direct comparison remains challenging as they were observed in separate measurements. Due to the similar experimental signatures of these effects a consistent assignment is difficult, leaving us with an incomplete picture. Here we present measurements on \textit{p-n} junctions in encapsulated graphene revealing several sets of magnetoconductance oscillations allowing for their direct comparison. We analysed them with respect to their charge carrier density, magnetic field, temperature and bias dependence in order to assign them to either snake states or Aharonov-Bohm oscillations.
Surprisingly, we find that snake states and  Aharonov-Bohm interferences can co-exist within a limited parameter range.
\end{abstract}

\maketitle

\section{Introduction}

Magnetoconductance effects, the change of the conductance as a function of magnetic field $B$, are both of fundamental significance (e.g. Aharonov-Bohm effect, Shubnikov-de-Haas oscillations \cite{IhnBook,NazarovBook}) and important for applications (e.g. GMR \cite{Baibich88,Binasch89}, TMR \cite{Julliere75}, etc.). Such effects have been investigated to a great extent also in two dimensional electron gases (2DEGs) realized in semiconductor heterostructures \cite{IhnBook,NazarovBook}. At low perpendicular magnetic fields electrons exhibit cyclotron motion that follows classical trajectories, allowing for the realization of electro-optical experiments such as transverse magnetic focusing \cite{Houten1989,Taychatanapat13}. At higher magnetic fields Landau levels are formed and electrons travel along edge channels \cite{IhnBook,NazarovBook}. Using electrostatic gating these channels can be guided within the sample, and using beam splitters based on quantum point contacts electronic Mach-Zehnder interferometers can be realized \cite{Ji2003, Samuelsson2004, Neder2007}. These interferometers enable the study of coherence effects of electronic states \cite{Bieri09, Litvin2007, Bocquillon2013}, noise in collision experiments \cite{Henny1999, Oliver1999} or probing the exotic nature of certain quantum Hall channels \cite{Bid2009, PhysRevLett.97.186803}.

Graphene not only offers similarly high mobility as 2DEGs, but it also allows for 0the formation of gapless \textit{p-n} interfaces not possible in conventional 2DEGs. Graphene \textit{p-n} junctions host quasi-classical snake trajectories at low field \cite{Williams11_2,Barbier12,Milovanovic14,Rickhaus15,Taychatanapat15,Kolasinski17}, where electrons curve back and forth along the opposite side of the \textit{p-n} junction. At high field edge channels propagate along the junction and coupling between these channels result in a Mach-Zehnder interferometer displaying the Aharonov-Bohm effect \cite{Morikawa15,Wei17}. Both effects result in magneto-conductance oscillations as a function of magnetic field and gate voltage. However, their similar signatures make it difficult to distinguish the two from each other. Moreover, experiments are performed within the transition between the classical and the quantum regime. Note that the observation of an Aharonov-Bohm effect requires phase coherent transport, while snake states are based on ballistic transport. Finally, Coulomb interaction of charge carriers localized in conducting islands coupled to edge channels can also lead to magnetoconductance oscillations \cite{Zhang09_3, Ilani2004, Halperin2011,C6NR00187D}.

Here we present measurements on high-mobility encapsulated graphene \textit{p-n} junctions, where several sets of magneto-conductance oscillations are observed simultaneously. Related oscillations have been observed \cite{Rickhaus15, Taychatanapat15,Morikawa15, Wei17}, but there is still an ongoing discussion on their origin. Their simultaneous observation, which we report here, allows a direct comparison with respect to the gate, field, temperature and bias dependence, resulting in a consistent and comprehensive assignment of the different oscillations.\\

The paper is organized as follows: First we introduce the most relevant concepts of snake states and Aharonov-Bohm oscillations along graphene \textit{p-n} junctions. Then we present measurements of several sets of oscillations within the bipolar regime. These magnetoconductance oscillations are carefully analysed with respect to their gate, magnetic field, temperature and bias dependence. We show that these oscillations can be attributed to either snake states or Aharonov-Bohm oscillations as introduced previously. We furthermore support our findings with theoretical models and quantum transport simulations.
Finally, we briefly discuss an additional type of magnetoconductance oscillation that has not been reported before.

\subsection{Snake states}

In small magnetic fields electrons follow skipping trajectories which turn to snake states along the \textit{p-n} junction \cite{Barbier12,Milovanovic14,Rickhaus15,Taychatanapat15}.
\begin{figure}[tb]
    \centering
      \includegraphics[width=1\columnwidth]{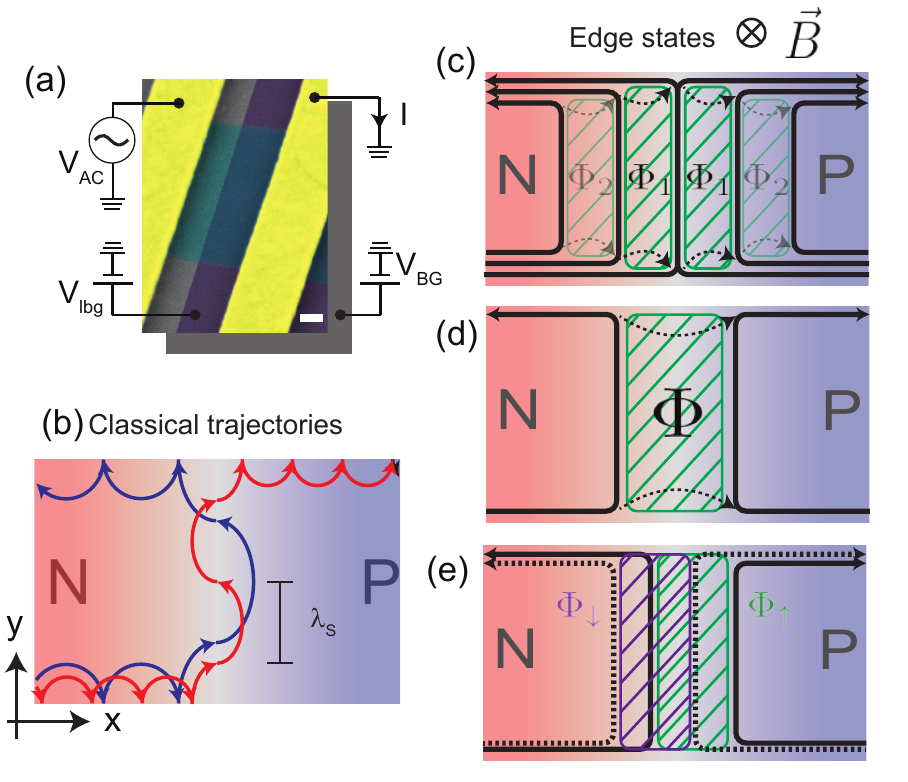}
    \caption{\textbf{Concept of snake states and Aharonov-Bohm interference along a graphene \textit{p-n} junction}. \textbf{a,} False-color SEM image of the device where the leads are colored yellow, the graphene encapsulated in hBN is colored cyan and the local bottom-gate (a structured few-layer graphite electrode underneath the hBN-graphene-hBN stack) is colored purple . Scale-bar equals $200\,$nm. $V_{BG}$ denotes the voltage applied to the global back-gate and $V_{lbg}$ is applied to a graphite bottomgates. \textbf{b,} Snake states seen in the framework of classical skipping orbits for two different magnetic field values (blue and red trajectories, $B_{red}>B_{blue}$). \textbf{c,} Principle of Aharonov-Bohm interference between quantum Hall edge-states propagating along the \textit{p-n} interface. At high bulk filling factors ($\nu_{L/R}$) several different areas are enclosed due to inter-channel scattering at the flake edges (green shaded area, and dashed arrows). Of these, the one that involves the least number of scattering events is expected to dominate ($\si{\Phi_1}$). \textbf{d,} At high magnetic fields Aharonov-Bohm interference can occur between the spatially separated edge states of the degeneracy-lifted lowest Landau level. The green area corresponds to the insulating region with local $\nu$ of 0. \textbf{e,} At even larger magnetic fields full degeneracy lifting occurs, and two spin-polarized interferometers are formed: purple area for spin-down (dashed) channels, green for spin-up (solid) channels. The interferometers are independent, as scattering between them is not allowed, since the spin is conserved along the edges.}
    \label{fig:Concept}
\end{figure}
These trajectories bend in opposite direction on the two sides of the \textit{p-n} junction due to an opposite Lorentz force, as sketched in Fig.~\ref{fig:Concept}b. Charge carriers with trajectories having a small incident angle with respect to the \textit{p-n} junction normal are transmitted very effectively from the n- to p-doped region of the graphene device (and vice versa) due to Klein-tunneling \cite{Klein29,Cheianov06,Kartnelson06}. In the  simplest case the \textit{p-n} junction is step-like and symmetric, and the cyclotron radius, $R_\text{C}=\lambda_S/2=\hbar k_\text{F}/(e B)$, is the same constant value on both sides. Here $k_F$ is the momentum of the electrons, $B$ the magnetic field and $\lambda_S$ is the size of the snake period or "skipping length". By changing B or the electron density n, and thus the cyclotron radius, the charge carriers end up either on the left or right side of the \textit{p-n} junction, similar to what is shown in Fig.~\ref{fig:Concept}b. This results in a conductance oscillation, where the conductance is determined by how the cyclotron radius compares to the length of the \textit{p-n} junction.

A more realistic model includes a gradual change of the charge carrier density across the \textit{p-n} junction, which is illustrated in  Fig.~\ref{fig:Concept}b. For a \textit{p-n} junction parallel to the y-direction this gives rise to a position-dependent electric field $\vec{E}_\text{x}$ (which in the case of constant E-field would lead to the well known $\vec{E} \times \vec{B}$ drift velocity). By solving the semiclassical equations of motion for an idealized graphene \textit{p-n} junction where the charge carrier density changes linearly from $n_L$ to $n_R$ over a distance of $d_n$, the skipping-length $\lambda_\text{S}$ is given by (see Supporting information, SI):
\begin{equation}\label{eq:lambda_s}
\lambda _\text{S}=\left(\frac{\pi \hbar}{e B}\right)^2 \frac{|n_\text{L}-n_\text{R}|}{d_\text{n}}.
\end{equation}
 Note that $S=|n_\text{L}-n_\text{R}|/d_\text{n}$ corresponds to the slope of the charge carrier density profile.
The conductance oscillations which can be measured across the \textit{p-n} junction of width $W$ at a given Fermi-energy $E$ can be described by a phenomenological model according to:
\begin{equation}\label{eq:SS_conductance}
G(E) \sim \cos \left( \pi \frac{W}{\lambda_\text{S}} \right),
\end{equation}
which describes the commensurability between $\lambda_\text{S}$ and $W$. The cosine itself accounts for a smooth conductance oscillation. Details of this model will be discussed later. We emphasize again that phase coherence is not required for this effect to appear.

\subsection{Aharonov-Bohm oscillations}\label{sec:AB_oscillations_theory}
While at low magnetic fields the motion of the charge carriers is well described using the picture of skipping and snake trajectories along edges and \textit{p-n} junctions, upon increasing the magnetic field one enters the quantum regime where transport is commonly described by edge states. The concept of interference formed by spatially separated edge states has already extensively been studied in 2DEGs, including the realization of Fabry-P\'{e}rot \cite{Chamon97}  and Mach-Zehnder \cite{Ji03} interferometers, while in graphene \textit{p-n} junctions it was first introduced by Morikawa et \textit{al} \cite{Morikawa15}. Here, edge states propagate on either side of the \textit{p-n} junction, and coupling between them is enabled at the junction's ends due to scattering on disordered graphene edges as illustrated in Fig.~\ref{fig:Concept}c,d.  Coupling between the edge states across the \textit{p-n} junction, illustrated in Fig.~\ref{fig:Concept}c by the black, dashed arrows, is restricted to the disordered graphene edges \cite{Morikawa15,Wei17}.
As the edge states encircle an enclosed area $A$ at finite perpendicular magnetic field $B$, the acquired Aharonov-Bohm phase is the magnetic flux, $\si{\Phi}=A B$. The conductance oscillations can be described phenomenologically:
\begin{equation}\label{eq:AB_conductance}
G(E) \sim \cos \left(2 \pi \frac{\si{\Phi}}{\si{\Phi}_0}\right),
\end{equation}
where $\si{\Phi}_0=$\si{h/e} is the magnetic flux quantum \cite{Aharonov59}.
In contrast to snake states, this is a phase coherent effect.

If multiple Landau levels are populated, several different interferometer loops, enclosing different areas, can contribute.
However, for the measured conductance across the \textit{p-n} junction only paths that connect the n- to the p-side are relevant. Of these, the ones with the least number of scattering events are expected to dominate the oscillation. These are the two inner ones denoted with $\si{\Phi}_1$ in Fig.~\ref{fig:Concept}c. The interference signal involves only one scattering event along each path, while for loops of type $\si{\Phi}_2$ at least two scattering events are necessary per path.

At high magnetic fields the Landau levels, which have valley and spin degeneracy at low field, can be partially (or fully) split \cite{Zhang06, Young12}. This leads to a spatial separation of the edge states associated with the lowest Landau level by an insulating region ($\nu=0$), as shown in Fig.~\ref{fig:Concept}d. Here, the valley degeneracy is lifted so that the edge state is still spin degenerate.

The idea of an Aharonov-Bohm interference, put forward in Ref.~\cite{Morikawa15}, was generalized by Wei et \textit{al}. \cite{Wei17} by considering full degeneracy lifting of the Landau levels, both in spin and valley. It was shown that the edges can mix the valleys, but not the spins, as sketched in Fig.~\ref{fig:Concept}e. Therefore scattering between edge states is only possible if they are of identical spin orientation. This gives rise to two sets of magnetoconductance oscillations - one for each spin-channel for bulk filling factors $|\nu_{L/R}|>2$ as described in Ref. \cite{Wei17}.
An increase of the spacing between neighbouring edge states is expected to decrease the scattering rate at the flake edge between edge states, giving rise to a reduced oscillation amplitude. At the same time the magnetic field needed to change the flux by a flux quantum is reduced, which will lead to changing magnetic field spacing. Details of the magnetic field spacing and the temperature dependence of the Aharonov-Bohm will be discussed later.

\begin{figure*}[tb]
    \centering
      \includegraphics[width=2\columnwidth]{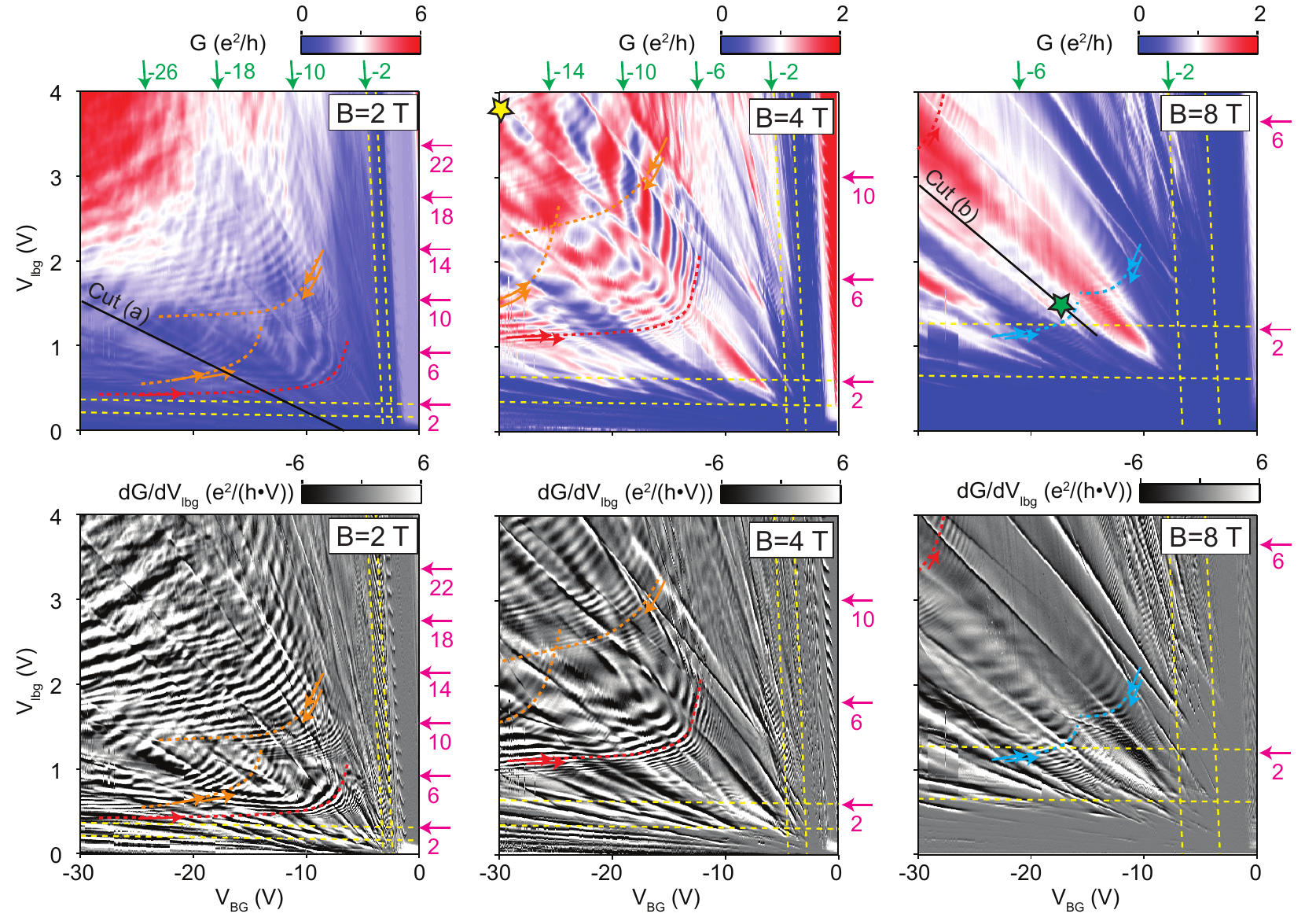}
    \caption{\textbf{Conductance (top panels) and its numerical derivative (bottom panels) of a \textit{p-n} junction in the bipolar regime for different magnetic fields.} The filling factors, obtained from a parallel-plate capacitor model, are given in green for the cavity tuned by the global back-gate ($\nu_{\text{BG}}$), and in purple for the cavity tuned by the local bottom-gate ($\nu_{\text{lbg}}$). The yellow, dashed lines indicate filling factors 1 and 2. The different types of magnetoconductance oscillations are indicated with the red, orange and cyan arrows/dashed curves. The lines indicate where the magnetic field dependencies of Fig.~\ref{fig:Gate_Field} were taken, whereas the stars indicate the position of the bias dependent measurements of Fig.~\ref{fig:Bias_spectroscopy}.}
    \label{fig:Gate_Gate}
\end{figure*}

\section{Measurements}
The hBN/graphene/hBN heterostructures were assembled following the dry pick-up technique described in Ref.~\cite{Wang13}. The full heterostructure was transferred onto a pre-patterned piece of few-layer graphene used as a local bottom-gate. Standard e-beam lithography was used to define the Cr/Au side-contacts, with the bottom hBN layer (\SI{70}{nm} in thickness) not fully etched through in order to avoid shorting the leads to the bottom-gates.
The graphene samples were shaped into \SI{1.5}{\micro m} wide channels using a CHF$_{3}$/O$_{2}$ plasma. A false-color SEM image of the final device is shown in Fig.~\ref{fig:Concept}a (for more details see SI).
The charge carrier mobility $\mu$ was extracted from field effect measurements yielding $\mu$\SI{\sim 80000}{cm^2 V^{-1} s^{-1}}.
The \textit{p-n} junction is formed by a global back- and a local bottom-gate which allows for independent tuning of the doping on each side of the \textit{p-n} junction. The presence of Fabry-P\'{e}rot oscillations (see SI) also attests to the high quality of our device. We have observed the magnetoconductance oscillations on $\sim10$ \textit{p-n} junctions in two separate stacks. Measurements were performed in a variable temperature insert with a base-temperature of $T=$\SI{1.5}{K} and a He-3 cryostat with a base-temperature of $T=$\SI{260}{mK}, using standard low-frequency lock-in techniques.

\subsection{Gate-gate dependence}
\begin{figure*}[tb]
    \centering
      \includegraphics[width=2\columnwidth]{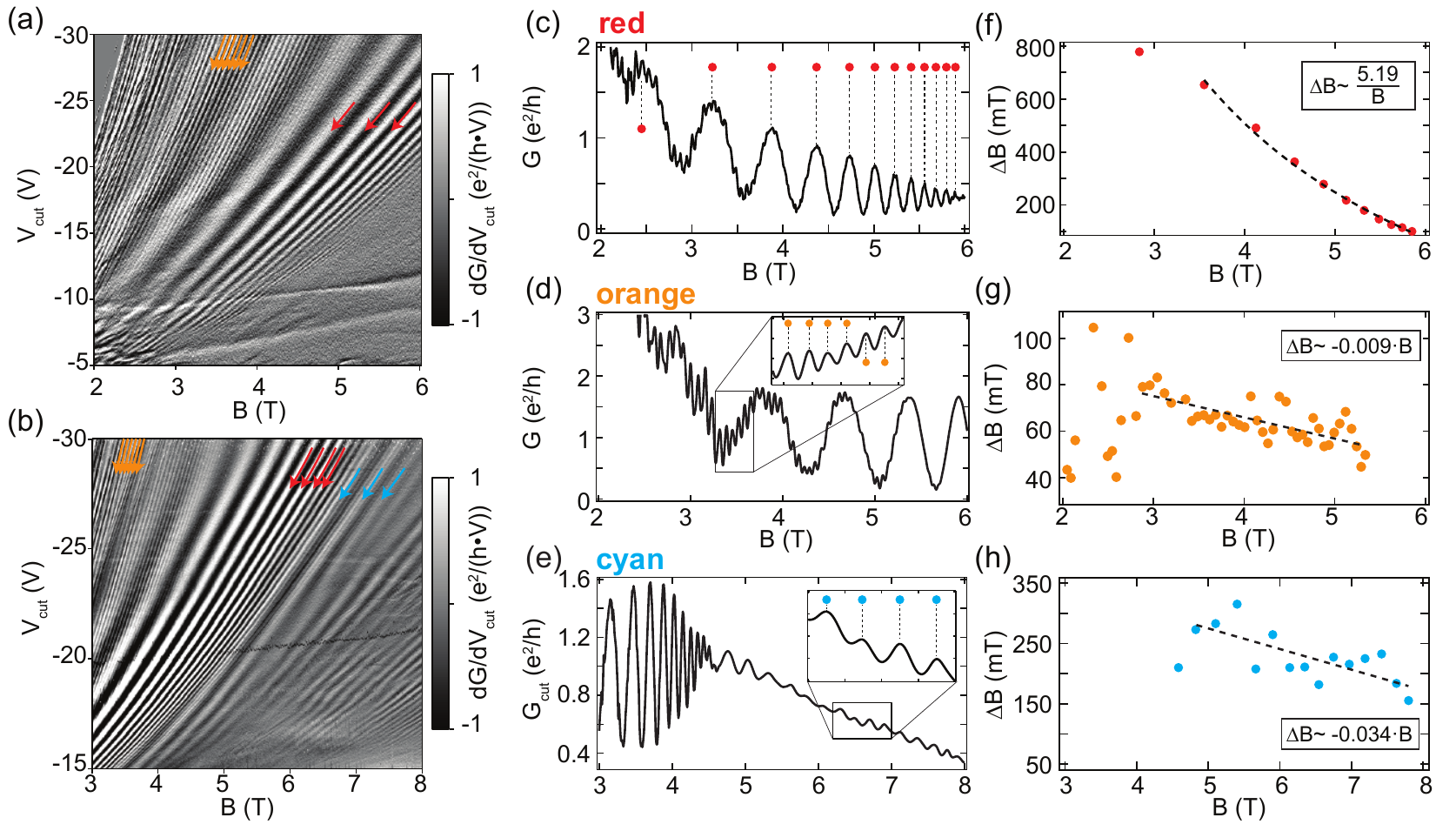}
    \caption{\textbf{Magnetic field dependence. a,} Numerical derivative of the conductance as a function of magnetic field and gate voltage as labelled in Fig.~\ref{fig:Gate_Gate} with ``Cut (a)''. Within a limited parameter range the magnetoconductance oscillations indicated with the red and orange arrows co-exist. The latter can be better seen in \textbf{b,} along the linecut labelled in Fig.~\ref{fig:Gate_Gate} with ``Cut (b)''. \textbf{c-e,} Conductance as a function of magnetic field for representative gate-gate configurations of the red ($V_\text{BG}=$\SI{-20}{V},$V_\text{lbg}=$\SI{1.8}{V}), orange ($V_\text{BG}=$\SI{-27.5}{V},$V_\text{lbg}=$\SI{4}{V}) and cyan ($V_\text{BG}=$\SI{-18.5}{V},$V_\text{lbg}=$\SI{1.27}{V}) magnetoconductance oscillations. The peak-positions are indicated with the red, orange and cyan dots. \textbf{f-h,} Magnetic field spacing between successive peaks ($\Delta B$) extracted from (c-e,). A $1/B$ and linear dependence of $\Delta B$ as a function of $B$ is indicated with the black dashed curve/lines for the snake states and Aharonov-Bohm interferences, respectively.}
    \label{fig:Gate_Field}
\end{figure*}
In Fig.~\ref{fig:Gate_Gate} the two-terminal conductance (top panels) and its numerical derivative (bottom panels) are shown as a function of the global back-gate ($V_\text{BG}$) and the local bottom-gate ($V_\text{lbg}$) within the bipolar regime at selected magnetic fields. Zero voltage of the global back-gate or local bottom-gate corresponds roughly to zero doping in the left or right side of the sample. In the gate-gate map, fine curved lines are visible along which the conductance is approximately constant, and perpendicular to these lines the conductance oscillates. Within the measured gate and field range we indentify three different types of magnetoconductance oscillations which are labelled with red, orange and cyan arrows/dashed lines. All of them have a roughly hyperbolic line shape being asymptotic with the zero-density lines related to either of the two sides of the samples. However, they are observed within different parameter ranges. The filling factors $\nu=n h /(e B)$, corresponding to the bulk values of the two sides tuned by the global back-gate and local bottom-gate, are indicated with the green and purple arrows in Fig.~\ref{fig:Gate_Gate}. The yellow dashed lines correspond to $|\nu |=1$ and $|\nu | =2$, for either side.

Upon comparing the different magnetoconductance oscillations it can be seen that the cyan ones exist at very low filling factors (starting at $|\nu|>1$), the red ones exist at intermediate filling factors $(\nu_\text{BG},\nu_\text{lbg})\sim (-4,4)$ and the orange ones appear at the highest filling factors. For one orange set, the filling factor values where the oscillations start to appear are around $(\nu_\text{BG},\nu_\text{lbg})\sim (-4,8)$, for the other orange set around $(\nu_\text{BG},\nu_\text{lbg})\sim (-8,4)$.
Furthermore, the spacing of neighbouring conductance oscillations as a function of charge carrier doping differs significantly for the cyan, red and orange oscillations.
An additional conductance modulation is also present where high and low conductance values follow lines that fan out linearly from the common charge neutrality point. The effect is more pronounced at higher magnetic fields and was attributed to valley-isospin oscillations \cite{Tworzydlo07, Low09, Qianfan2018} which are discussed in detail in Ref.~\cite{Handschin17}.\\
Whereas for the red magnetoconductance oscillations only one set is observed, two sets are observed for the orange and cyan ones. The latter ones are furthermore shifted in doping with respect to each other.
Using a device with a geometry enabling gate defined \textit{p-n-p} or \textit{n-p-n} junctions (see SI) we have excluded the possibility that the two orange sets of magnetoconductance oscillations originate from an additional \textit{p-n} junction formed between n-doped graphene near the Cr/Au contacts and a p-doped bulk. This is in agreement with quantum transport simulations (discussed later in this manuscript), which reproduce a double set of oscillations, in the same range where the orange ones are observed, without introducing contact doping. Therefore, a double set of oscillations must be the sign of two different interferometer loops working simultaneously near the \textit{p-n} junction in the bulk (see Fig.~\ref{fig:Concept}b).

\subsection{Magnetic field dependence}
Next we measured selected linecuts as indicated in Fig.~\ref{fig:Gate_Gate} with ``Cut (a)'' and ``Cut (b)'' as a function of magnetic field. The differential conductances as a function of magnetic field and gate voltage are shown in Fig.~\ref{fig:Gate_Field}a,b. The three magnetoconductance oscillations, which are labelled with the red, orange and cyan arrows, follow a roughly (but not exactly) parabolic magnetic field dependence where the oscillations shift to higher absolute gate voltages with increasing magnetic field. Furthermore, we observe a co-existence of multiple oscillations within a limited parameter range. The co-existence of the red and orange oscillations is seen in both Fig.~\ref{fig:Gate_Field}a and Fig.~\ref{fig:Gate_Field}b.
The conductance as a function of the magnetic field, while keeping the charge carrier densities on both sides of the \textit{p-n} junction fixed, is plotted in Fig.~\ref{fig:Gate_Field}c-e for three selected configurations.
In Fig.~\ref{fig:Gate_Field}c,d large oscillations (red in the previous graphs) with peak-to-peak amplitudes reaching nearly \SI{2}{e^2/h} can be seen. Within a limited parameter range there are smaller oscillations (orange in the previous graphs) superimposed on top of the red oscillations, having amplitudes reaching up to \SI{\sim 0.6}{e^2/h}. The cyan oscillations show amplitudes in the range $\sim0.05-0.1\,$e$^2/h$.
The magnetic field spacing ($\Delta B$) between neighbouring peaks is given in Fig.~\ref{fig:Gate_Field}f-h for the corresponding oscillations shown in Fig.~\ref{fig:Gate_Field}c-e. Even though all three types of magnetoconductance oscillations reveal a different spacing of $\Delta B$, they share a common trend, namely the decrease of $\Delta B$ with increasing $B$. Nevertheless, the rate of $\Delta B$ as a function of $B$ is quite different for the red compared to the orange and blue magnetoconductance oscillations, which is an indication that different physical mechanism are involved.

\subsection{Temperature dependence}
\begin{figure}[tb]
    \centering
      \includegraphics[width=1\columnwidth]{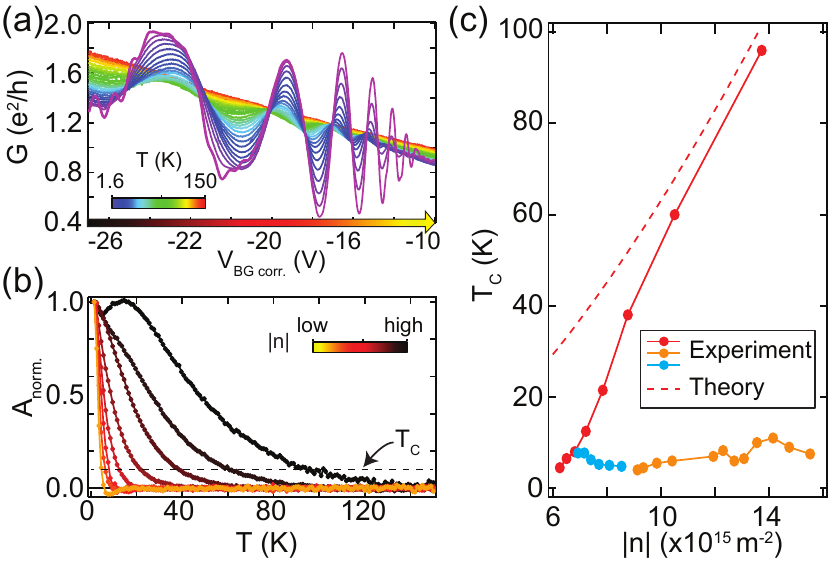}
    \caption{\textbf{Temperature dependence. a,} Red magnetoconductance oscillations as a function of the global back-gate ($V_\text{lbg}$ is chosen such that $|n_\text{BG}| \sim |n_\text{lbg}|$) and temperature at $B=$\SI{3.5}{T}. \textbf{b,} . $A_{\mathrm{norm}}$ of the red oscillations (the dominant ones in (a)) is plotted here as a function of temperature at various densities. The color coding corresponds to the x axis of panel (a). \textbf{c,} The solid lines/dots show the experimental values of $T_\text{C}$, defined as the temperature for which the oscillation amplitude is reduced to \SI{10}{\percent} of its low temperature value of the red, orange and cyan magnetoconductance oscillations (extracted at $B=$\SI{3.5}{T}, $B=$\SI{3}{T} and $B=$\SI{8}{T} respectively) as a function of charge carrier doping. The red, dashed line corresponds to the vanishing of snake states according to equation~\ref{eq:T_smearing} using $d_\text{n}=$\SI{50}{nm} and $W=$\SI{1500}{nm}.}
    \label{fig:Temp_dependence}
\end{figure}
In Fig.~\ref{fig:Temp_dependence} the temperature dependence of the red, orange and cyan magnetoconductance oscillations is given. Fig.~\ref{fig:Temp_dependence}a shows the red oscillations as a function of gate voltage and temperature ($|n_\text{BG}| \sim |n_\text{lbg}|$  and $B=$\SI{3.5}{T}). We characterize the temperature dependence of each oscillation by calculating the area $A$ under the oscillation with respect to the high-T smooth background. From this the normalized area, which is defined as $A_\text{norm.}=A(T)/A(T=1.6K)$, can be extracted at different densities, and is plotted as a function of temperature in Fig.~\ref{fig:Temp_dependence}b. A characteristic temperature for the disappearance of the oscillations, $T_c$, is then defined according to $A_{\textrm{norm}}(T_c)=0.1$. In Fig.~\ref{fig:Temp_dependence}c $T_\text{C}$ is plotted as a function of the density for all three types of magnetoconductance oscillations. While the red magnetoconductance oscillation reveals a significant temperature dependence as a function of the charge carrier density, surviving up to $T\sim$\SI{100}{K} at high doping, the orange and cyan magnetoconductance oscillations vanish at temperatures around $T\sim$\SI{10}{K} irrespective of the charge carrier density. This suggests again that different mechanisms are responsible for the red magnetoconductance oscillations compared to the orange and cyan magnetoconductance oscillations. Ballistic effects, such as snake states and transverse magnetic focusing, are known to survive to temperatures up to $T\sim$\SIrange{100}{150}{K} \cite{Taychatanapat13,Taychatanapat15,Lee16}. On the other hand, phase coherent transport in similar devices vanishes at temperatures around \SI{\sim 10}{K} (see Ref.~\cite{Zihlmann17_toPublish}).

\subsection{Bias dependence}

\begin{figure*}[tb]
    \centering
      \includegraphics[width=2\columnwidth]{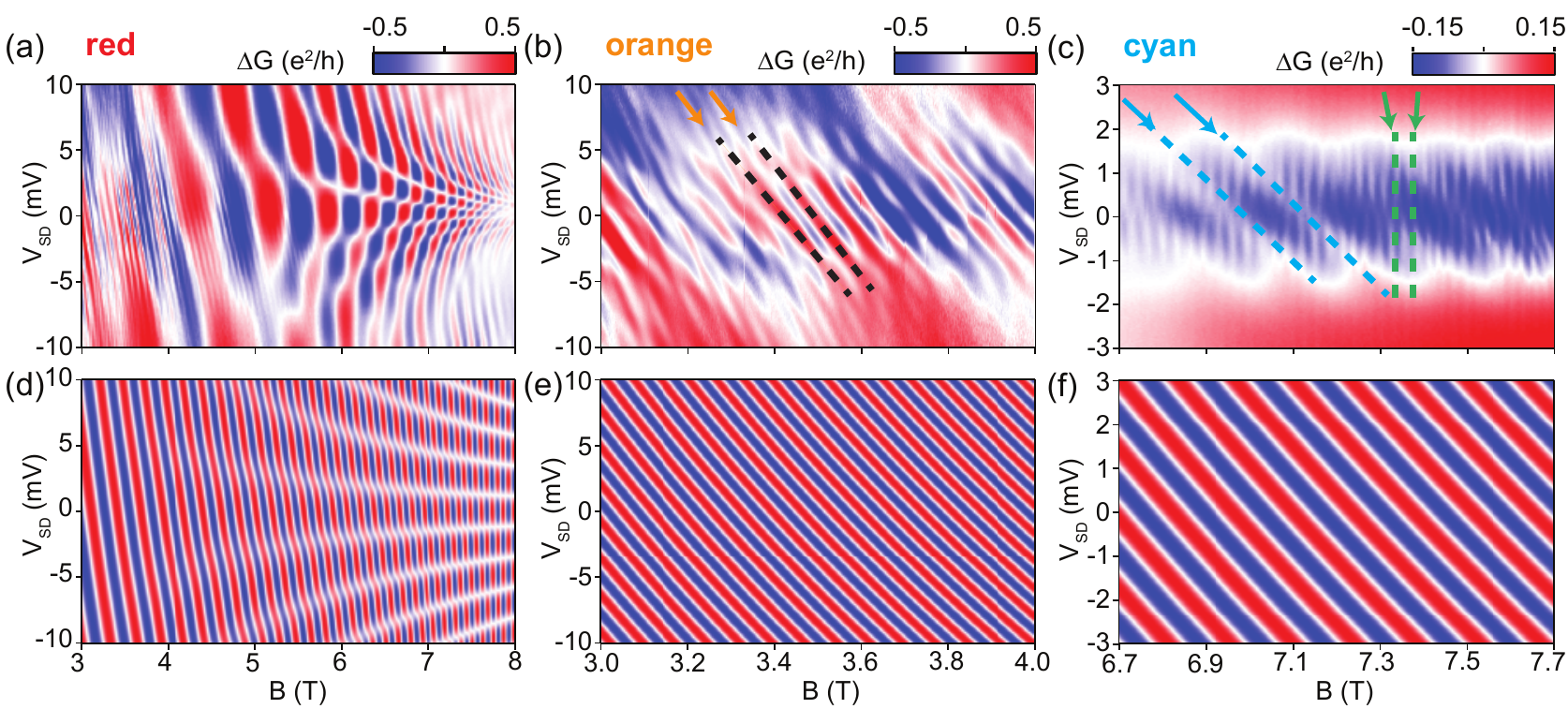}
    \caption{\textbf{Bias spectroscopy. a-c} Measurement of the red, orange and cyan magnetoconductance oscillations as a function of bias and magnetic field where a smooth background was subtracted. Gate-voltages remain fixed and are indicated in Fig.~\ref{fig:Gate_Gate} with the yellow (red and orange oscillations) and green (cyan oscillation) stars. \textbf{d-f,} Phenomenological simulations of the bias dependence of snake state and interference-induced oscillations(a-c).  Parameters used: $W=$\SI{1.5}{\micro m}, $d_\text{n}=$\SI{100}{nm} (red oscillations), $k_\text{F}$ corresponding to $n\sim$\SI{1.7e12}{cm^{-2}} (red, orange) or $n\sim$\SI{0.8e12}{cm^{-2}} (cyan). For the Aharonov-Bohm oscillations we considered a bias dependent gating effect with $\alpha=$\SI{0.32}{nm/mV_\text{SD}} and $d=$\SI{40}{nm} (orange oscillation) or $\alpha=$\SI{0.25}{nm/mV_\text{SD}} and $d=$\SI{20}{nm} (cyan oscillations), while a renormalization of the edge state velocity is neglected ($\beta=1$).}
    \label{fig:Bias_spectroscopy}
\end{figure*}

We have also investigated the  bias dependence of the different oscillations as a function of magnetic field while keeping the charge carrier densities fixed. The bias was applied asymmetrically at the source, while the drain remained grounded. The red magnetoconductance oscillations evolve from a tilted line pattern at smaller magnetic fields into a checker-board pattern at high magnetic field as shown in Fig.~\ref{fig:Bias_spectroscopy}a (a smooth background is subtracted). At high magnetic field the visibility of the checker-board pattern decreases with increasing $V_\text{SD}$ while a similar behaviour is absent (within the applied bias range of \SI{\pm 10}{mV}) for the tilted pattern. The bias dependence of the orange and cyan magnetoconductance oscillations is shown in Fig.~\ref{fig:Bias_spectroscopy}b,c, both revealing a tilted line pattern within the measured magnetic field range, as shown by the dashed lines and arrows. The bias dependence of the orange oscillations persists to \SI{\pm 10}{mV}, whereas that of the cyan oscillations vanishes around roughly \SI{\pm 2}{mV}.
In Fig.~\ref{fig:Bias_spectroscopy}c additional magnetoconductance oscillations with a narrow spacing of roughly $\Delta B \sim$\SIrange{4}{6}{mT} can be observed, indicated by green arrows and green dashed lines. These oscillations will be briefly discussed at the end of the paper.

\section{Discussion}
We have observed different magnetoconductance oscillations, marked with red, orange and blue. All of the oscillations have a roughly hyperbolic line shapes in the gate-gate map, but the magnetic field spacing, the temperature dependence and the fact that there is only a single set of red oscillations suggest that they are governed by different physical mechanisms.
Based on the experimental evidence presented until now, it is suggestive to assign the red oscillations to snake states and the others to the Aharonov-Bohm effect. This will be substantiated further on below.

\subsection{Magnetoconductance oscillations marked in red}
The red magnetoconductance oscillations start to appear in the range $|\nu| \sim 3-6$ as can be seen in Fig.~\ref{fig:Gate_Gate}. This corresponds to an occupation of roughly two edge states ($\nu=\pm 4$, Landau levels $0$ and $\pm 1$) without taking degeneracy lifting into account. 
The shape of the red magnetoconductance oscillations fits very well to what is expected for snake states following equation~\ref{eq:lambda_s} and equation~\ref{eq:SS_conductance}, as we will show below.

\begin{figure}[tb]
    \centering
      \includegraphics[width=1\columnwidth]{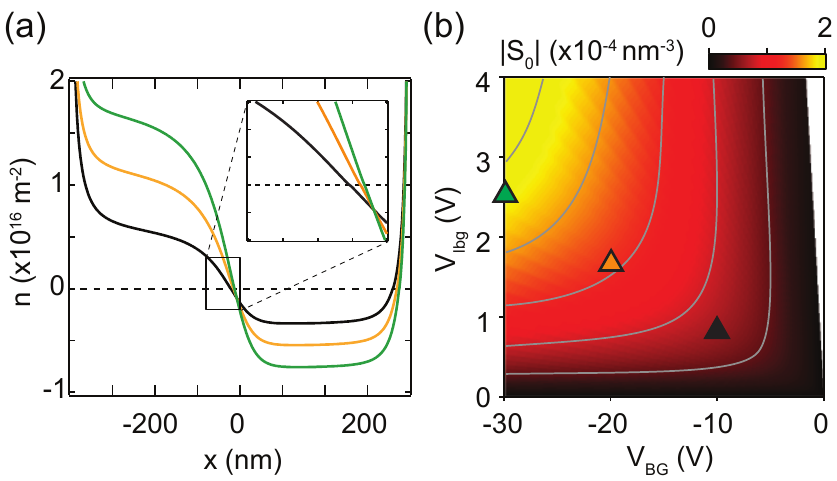}
    \caption{\textbf{Charge carrier density profile in the bipolar regime and extracted slope. a,} Representative charge carrier density profiles calculated from electrostatics at positions as indicated in (b) with the triangles. At $n=0$ the slope is nearly linear (inset). \textbf{b,} Slope $|S_0|$ extracted at $n=0$ as a function of the gates. Grey curves represent constant values of $|S_0|$, and consequently of $\lambda _\text{S}$, as well (equation~\ref{eq:lambda_s}).}
    \label{fig:Slope_Gate_gate}
\end{figure}

As discussed in the introduction, the oscillation results from a commensurability relation of the \textit{p-n} junction length and the skipping length, where the conductance is high or low depending on whether the snaking trajectories end up on the source or the drain side.
If the magnetic field is fixed, the skipping-length $\lambda_S$ is directly proportional to the slope of the \textit{p-n} junction according to equation~\ref{eq:lambda_s}. In Fig.~\ref{fig:Slope_Gate_gate}a the calculated charge carrier density profile at $B=$\SI{0}{T} is shown for three exemplary gate-gate configurations (details of the electrostatic simulations are given in the SI). It is clear that $S_0$ characterizes well the density profile in the vicinity of the \textit{p-n} junction. $S_0$ as a function of gate voltages is plotted in Fig.~\ref{fig:Slope_Gate_gate}b: here curves of constant $S_0$, and therefore of constant $\lambda _\text{S}$ (if $B$ remains fixed), follow a roughly hyperbolic line shape in agreement with the shape of the red magnetoconductance oscillations (Fig.~\ref{fig:Gate_Gate}). Although the orange and cyan oscillations also seem to follow hyperbolic shapes on gate-gate maps, the red ones only have a single set as expected of snake states.

Next we analyze the magnetic field dependence and spacing expected of snake state induced oscillations. Based on $S_0$ (Fig.~\ref{fig:Slope_Gate_gate}b) one can calculate the conductance contribution as a function of an arbitrary linecut and magnetic field (not shown here), leading to a roughly parabolic magnetoconductance oscillation which strongly resembles the measurements shown in Fig.~\ref{fig:Gate_Field}a,b.\\
By using the model with a constant density gradient the magnetic field spacing as a function of magnetic field is given approximately by (see SI):
\begin{equation}\label{eq:SS_B_spacing}
\Delta B \sim 2\frac{\pi ^2 \hbar ^2 n}{e^2 W d _\text{n}}\frac{1}{B},
\end{equation}
where a symmetric \textit{p-n} junction with $n\equiv |n _\text{L}|=|n _\text{R}|$ was assumed. The magnetic field spacing in the experiment is very well described by the $1/B$ dependence. By fitting the magnetic field spacing of the snake state model as described in equation~\ref{eq:SS_B_spacing} to the measurements shown in Fig.~\ref{fig:Gate_Field}f we extracted a slope of $S=$\SI{1.82e-3}{nm^{-3}} which is roughly one order of magnitude larger than what was calculated in Fig.~\ref{fig:Slope_Gate_gate}b ($S_0 \sim$\SI{1.2e-4}{nm^{-3}}). One explanation for the discrepancy is that strictly speaking $S_0$ is only valid at $B=$\SI{0}{T}. However, at finite magnetic field the charge carrier density has to be calculated self-consistently leading to areas with a constant charge carrier density (compressible region) and areas where the charge carrier density changes rapidly ($S>S_0$, incompressible regions) \cite{Chklovskii92}. Also, the model supposes that the trajectories stay within the area with a constant density slope (see SI), which might be not valid at low fields due to the increased cyclotron radius and skipping length.

The decrease of the oscillation amplitude with increasing magnetic field (Fig.~\ref{fig:Gate_Field}c) is compatible with the picture of classical snake trajectories, where the conductance oscillation results from the sum over all trajectories which form caustics along the \textit{p-n} junction \cite{Davies12,Patel12}. Upon increasing the magnetic field the charge carriers have to pass the \textit{p-n} junction more often (decreasing $\lambda_\text{S}$). This leads to a reduced oscillation amplitude \cite{Kolasinski17} because only trajectories with an incident angles being perpendicular to the \textit{p-n} junction ($\theta=0$) have a transmission probability of $t=1$, while for all remaining trajectories $t < 1$ is valid \cite{Kartnelson06,Cheianov06,Chen16}.\\
Our most compelling argument for identifying the red oscillations with snake states comes from the comparison of the measured temperature dependence with that calculated by the following simple model. At finite temperatures $T$ the Fermi-surface is broadened by $\Delta E\sim k_\text{B} T$ (where $k_\text{B}$ is the Boltzmann constant), thus leading to a spread of the Fermi-wavevector according to $\Delta k_\text{F}  \sim k_\text{B} T/(\hbar v_\text{F})$. The oscillations are expected to vanish if the smearing of trajectories becomes comparable to half a period:
\begin{equation}\label{eq:T_smearing_start}
2 \left( \lambda_\text{S,max}-\lambda_\text{S,min}\right) \cdot N \sim \left< \lambda_\text{S} \right>,
\end{equation}
where $\lambda_\text{S,max}$, $\lambda_\text{S,min}$ and $\left< \lambda_\text{S} \right>$ correspond to the maximal, minimal and average skipping-length, respectively and N to the number of skipping periods. This leads to a characteristic temperature

\begin{equation}\label{eq:T_smearing}
T_c \approx \frac{2 v_\text{F} \hbar^3}{W d_\text{n} k_\text{B} e^2 B^2} \sqrt{n^3 \pi ^5},
\end{equation}
where the oscillations vanish. Here $k_\text{B}$ is the Boltzmann constant. Details of the calculation can be found in the SI. The vanishing of the red magnetoconductance oscillations with increasing charge carrier doping, which is plotted in Fig.~\ref{fig:Temp_dependence}c (red, dashed line), is in good agreement with what is expected for snake states according to Equation~\ref{eq:T_smearing}, unlike the other type of oscillations.

Finally, we analyze the bias dependence of snake states. Details of the model can be found in the SI. The bias dependence is calculated by taking into account the energy dependence of the snake-period through its momentum dependence. In the case of a fully asymmetric bias the model
reproduces the tilted pattern which is shown in Fig.~\ref{fig:Bias_spectroscopy}d at low magnetic field. On the other hand, for the case of completely symmetric bias, the same model leads to the checker-board pattern which is shown in Fig.~\ref{fig:Bias_spectroscopy}d at high magnetic field. The checker-board pattern is in agreement with previous studies \cite{Morikawa15, Wei17}, where a similar behaviour was observed. The oscillation period decreases with increasing magnetic field in the simulation (Fig.~\ref{fig:Bias_spectroscopy}d) comparable to the experiment (Fig.~\ref{fig:Bias_spectroscopy}a). In order to reproduce the transition from tilted (asymmetric biasing) to checker-board pattern (symmetric biasing) we varied the bias asymmetry going from low to high magnetic field. We speculate that it might be related to the capacitances in the system \cite{McClure09}, but the precise reason remains unknown so far. We discuss this in more detail in the SI.

\subsection{Magnetoconductance oscillations marked in orange}
From all the observed magnetoconductance oscillations the orange ones occur at the highest filling factors starting at roughly $|\nu| \sim 6$ and persisting up to $|\nu| = 20$ or even higher, as shown in Fig.~\ref{fig:Gate_Gate}. This corresponds to an occupation of at least two edge states ($|\nu|=0$ and $|\nu|=4$) without taking a possible degeneracy lifting into account. Snake states can be excluded here, as the double set of oscillations indicates two simultaneous effects, with their origin displaced in real space with respect to the \textit{p-n} junction. Therefore, we attribute these oscillations to Aharonov-Bohm oscillations between quantum Hall edge states propagating in parallel with each other and the \textit{p-n} junction. Their temperature, B-field and bias dependence also support this idea. Below we discuss expectations and compare them with our experimental observations. \\
In an Aharonov-Bohm interferometer, the magnetic field spacing $\Delta B$ between neighbouring conductance peaks is given by:
\begin{equation}\label{eq:AB_B_spacing}
\Delta B=\frac{h}{e}\frac{1}{A}.
\end{equation}
Here $\si{\Phi}_0=$\si{h/e} is the magnetic flux quantum and $A$ is the enclosed area given by the product of the width of the flake $W$ and the distance of the edge states, $d$.

This suggests a constant $\Delta B$ for a fixed spacing $d$.
However, in the experiments $\Delta B$ is not exactly constant because the real-space positions of the edge states, which define $A$, vary as a function of magnetic field and the \textit{p-n} junction's density profile \cite{Morikawa15}. By considering a linear charge carrier density profile $\Delta B$ decreases linearly with increasing $B$.
This is in agreement with what was measured in Fig.~\ref{fig:Gate_Field}g,h, indicated with the black dashed line, therefore suggesting an Aharonov-Bohm type of interference.
Even though multiple areas might be enclosed between the various edge states, only one Aharonov-Bohm loop will dominate as explained previously and sketched in Fig.~\ref{fig:Concept}b. The magnetic field spacing of the orange magnetoconductance oscillations (Fig.~\ref{fig:Gate_Field}g) was converted into a distance ranging from $d$\SI{\sim 30}{nm} at $B$\SI{\sim 2}{T} to $d$\SI{\sim 55}{nm} at $B$\SI{\sim 5.5}{T}. The decreasing oscillation amplitude ($\Delta G_\text{osc}$) with increasing magnetic field (Fig.~\ref{fig:Gate_Field}d) directly indicates the vanishing coupling between edge states as they move further apart from each other at higher magnetic fields.\\
We have used the zero-field electrostatic density profile shown in Fig.~\ref{fig:Slope_Gate_gate}a to identify the spacing $d$ between two edge state for any set of ($V_\text{BG}$,$V_\text{lbg}$) within the gate-gate map. The magnetoconductance oscillation can then be calculated according to equation~\ref{eq:AB_conductance}, leading to a roughly hyperbolic shape as a function of the two gates at fixed magnetic field (see Supporing Informations). The two sets of the orange oscillations can be reproduced with a double Aharonov-Bohm interferometer as sketched in Fig.~\ref{fig:Concept}b, where the conductance oscillations arising from the interferometer on the left (e.g. quantum Hall channel with $\nu=0$ and $\nu=\pm 4$) and right (e.g. $\nu=0$ and $\nu=\mp 4$) side are added up incoherently. The two sets of orange magnetoconductance oscillations are slightly shifted in doping with respect to each other because each of the two gates tunes one side of the \textit{p-n} junction more effectively.
Furthermore, measuring a linecut as a function of magnetic field reveals a roughly parabolic trend (see SI). These findings are in good agreement with the measurements which are shown in Fig.~\ref{fig:Gate_Gate} and Fig.~\ref{fig:Gate_Field}a,b, respectively.\\
In interference experiments which depend on phase coherent transport, a vanishing of the oscillation pattern with temperature can have different origins such as loss of phase coherence due to enhanced inelastic scattering events. As soon as $l_{\mathrm{\Phi}} < L$, where $l_{\mathrm{\Phi}}$ is the phase coherence length and $L$ is the total path length, the interference pattern is almost completely lost. As mentioned before, the phase coherence length is below $1-2\,\mu$m in similar devices at temperatures around \SI{\sim 10}{K} (see Ref.~\cite{Zihlmann17_toPublish}). However, the interference can as well be lost at finite temperatures even if $l_{\mathrm{\Phi}} > L$ if the two interfering paths have different lengths ($\Delta L \neq 0$), again due to the smearing of the Fermi wavevector.  In this case, the interference pattern is expected to vanish at temperatures around
\begin{equation}\label{eq:Temp_dep_coherence}
T=\frac{h v_{\text{F}}}{k_{\text{B}} \Delta L}.
\end{equation}

Since for the Aharonov-Bohm interference along a graphene \textit{p-n} junction $\Delta L$ is ideally zero (see Fig.~\ref{fig:Concept}b,c) or very small, this effect is negligible. Consequently, the loss of the interference signal with increasing temperature depends on the decrease of $l_{\mathrm{\Phi}}$, which depends only weakly on the charge carrier doping \cite{Zihlmann17_toPublish}, in agreement with Fig.~\ref{fig:Temp_dependence}c.

Finally, we calculate the bias dependence of Aharonov-Bohm oscillations. The details of the model are discussed in the SI.
The bias dependence is introduced via the momentum difference which leads to:
\begin{equation}\label{eq:AB_bias_1}
G(V) \sim \cos \left[2 \pi\frac{W \cdot (d + \alpha V) \cdot B}{\Phi_0} + k \Delta L \right].
\end{equation}
Here $\alpha$ is a phenomenological parameter in order to account for a bias dependent gating effect \cite{Bieri09}. For simplicity the edge state spacing $d$ is modified proportional to the applied bias. The factor $k \Delta L$ in Equation~\ref{eq:AB_bias_1} accounts for a possible path-difference between the edge states, where $k$ is replaced by $k=k_\text{F}+ eV/(\hbar v_\text{F} \beta)$. The parameter  $\beta$ ($0\leq \beta \leq 1$) was introduced to account for the renormalized edge state velocity compared to the Fermi velocity \cite{Cohnitz16}.

We have found that the second term alone ($k\Delta L$ at $\alpha=0$) cannot lead to substantial bias dependence if the parameters are chosen realistically ($\Delta L\sim$\SI{20}{nm} and $\beta =1$). To account for the tilt of the measurement a considerable renormalization of the edge state velocity is needed leading to an unphysically large reduction of $v_F$ by a factor of one hundred. Therefore most of the tilt must come from non-zero $\alpha$ and the bias induced gating effect. \\

The applied bias voltage shifts the electrochemical potential on one side (or both sides) and therefore leads to a change of the density profile. The changing density profile results in shifting of the edge states, and in order to keep the flux through the interferometer fixed, the magnetic field has to be changed. Assuming that the applied bias affects the edge state spacing according to $\Delta d=\alpha \cdot$\si{ V_\text{SD}}, then $\alpha$ can be extracted from the bias spacing in Fig.~\ref{fig:Bias_spectroscopy}b. This leads to values of $\alpha \sim$\SI{0.32}{nm/mV_\text{SD}}. The resulting bias dependence is plotted in Fig.~\ref{fig:Bias_spectroscopy}e. Based on a simple model, it is possible to numerically estimate the value of alpha, to compare with the observation. We keep the width $d_n$ of the \textit{p-n} junction constant, and take the bias voltage into account directly changing the left and right band offset and thus $n_{L/R}$.  From this simple model we obtain values in the order of $\alpha\sim$\SIrange{0.39}{0.48}{nm/mV_\text{SD}} for the orange magnetoconductance oscillations, which agrees fairly well with the experimental observations. Details are given in the SI.

\subsection{Magnetoconductance oscillations marked in cyan}
The cyan magnetoconductance oscillations were observed at the lowest filling factors as low as $|\nu|\sim 2$ or even less, above $B\sim$\SI{4}{T} as shown in Fig.~\ref{fig:Gate_Gate}. We attribute these oscillations to Aharonov-Bohm oscillations formed by edge channels of the fully degeneracy lifted lowest Landau level, as shown in Fig.~\ref{fig:Concept}e). Since full degeneracy lifting of the lowest Landau level (valley and spin) is observed for $B>$\SI{5}{T} (see SI), the edge states are spin and valley polarized. While the spin degree of freedom is conserved along the edges of graphene and along the \textit{p-n} junction \cite{Wei17}, the valley degree of freedom is only conserved along the \textit{p-n} junction. Mixing between the edge states of the lowest Landau levels having equal spin is consequently prohibited along the \textit{p-n} junction, but possible at the graphene edges.
Comparable to the orange magnetoconductance oscillations, the magnetic field spacing of the cyan oscillations decreases monotonically, corresponding to an edge state spacing of $d\sim$\SI{9}{nm} at $B=$\SI{4.5}{T} and to $d\sim$\SI{15}{nm} at $B=$\SI{8}{T}. 
The oscillation amplitude of the cyan magnetoconductance oscillation shown in Fig.~\ref{fig:Gate_Field}c was rather constant with magnetic field including some irregularities. We note that the cyan magnetoconductance oscillations are predominantly visible at a charge carrier doping of $|n_\text{BG}|\sim |n_\text{lbg}|$ (Fig.~\ref{fig:Gate_Gate}) for reasons which are unknown yet.
Similar to the orange magnetoconductance oscillations, the temperature dependence of the cyan ones depends only slightly on the charge carrier density, and is most likely related to the loss of phase coherence.

Finally, the bias dependence is modelled similarly to the orange oscillation. The measurements can be well reproduced by using $\alpha \sim$\SI{0.25}{nm/mV_\text{SD}} to account for the bias dependent gating effect as demonstrated in Fig.~\ref{fig:Bias_spectroscopy}f, which shows good agreement with our measurements (panel c). Our simple estimate using the model detailed in the SI gives values in the order of $\alpha\sim$\SIrange{0.16}{0.25}{nm/mV_\text{SD}} for the cyan magnetoconductance oscillations, again agreeing fairly well with the experimental findings.

\section{Quantum transport simulations}
\begin{figure}[tb]
    \centering
      \includegraphics[width=1\columnwidth]{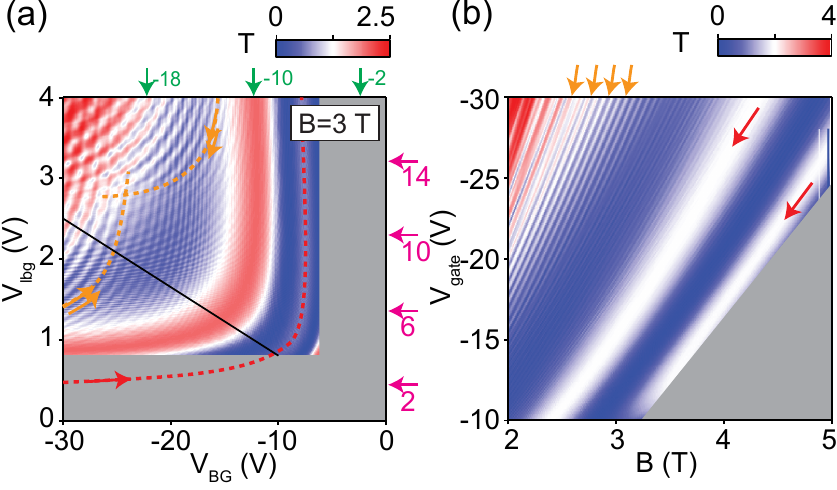}
    \caption{\textbf{Quantum transport calculations for a graphene \textit{p-n} junction in magnetic field. a,} Transmission function ($T$) of charge carriers through the \textit{p-n} junction with the same gate geometry as measured one, as a function of a local bottom-gate and a global back-gate. Red and orange magnetoconductance oscillations are indicated with the dashed curves/arrows. Filling factors of the global back-gate and the local bottom-gate are indicated with the green/purple arrows. Low doping values (shaded in grey) were omitted to reduce the computational load. \textbf{b,} Linecut as indicated in (a) with the black line as a function of magnetic field.}
    \label{fig:Quantum_Transport_Simulation}
\end{figure}
To complement our measurements, we additionally performed quantum transport calculations based on so-called scaled graphene \cite{Liu15}
using the realistic device geometry. These calculations were able to reproduce the red and orange magnetoconductance oscillations. In the calculations electron-electron interactions is not taken into account. In Fig.~\ref{fig:Quantum_Transport_Simulation}a the conductance is shown as a function of the local bottom-gate and the global back-gate at $B=$\SI{3}{T} within the bipolar regime. Comparable to the measurements presented in Fig.~\ref{fig:Gate_Gate}, two sets of magnetoconductance oscillations can be seen which are shifted in doping, which we identify with the orange ones. A few ridges also appear which we assign to the red oscillations. In Fig.~\ref{fig:Quantum_Transport_Simulation}b the evolution of the red and orange oscillations are shown as a function of gate  and magnetic field. The calculations show that the orange magnetoconductance oscillations seen in the experiments can be reproduced without the splitting of the lowest Landau level in contradiction with the claims of Ref.~\cite{Wei17}. In that work all oscillations were linked to the split lowest Landau levels, however our analysis shows that only the blue oscillations, appearing at the lowest filling factors, can be attributed to degeneracy lifted Landau levels.

\section{Additional magnetoconductance oscillations at high magnetic field}

\begin{figure}[tb]
    \centering
      \includegraphics[width=1\columnwidth]{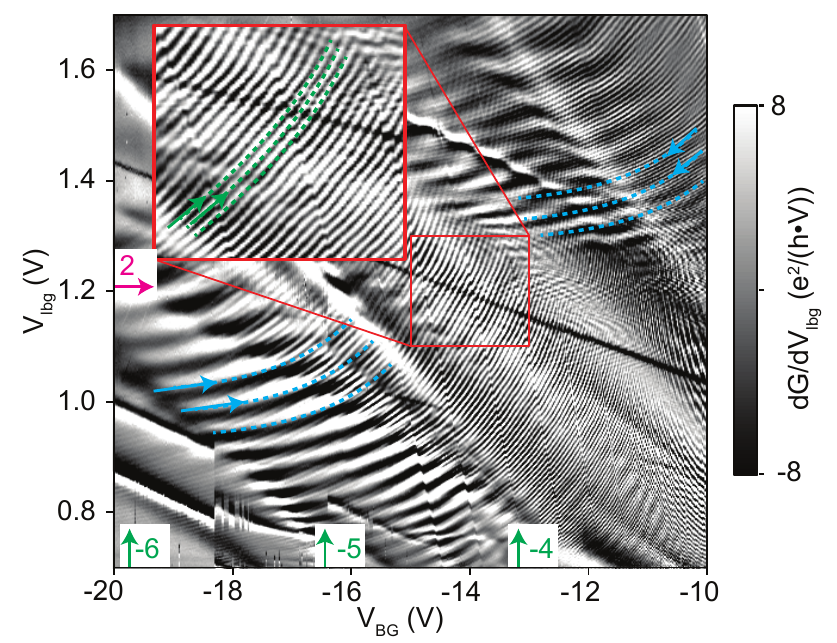}
    \caption{\textbf{Additional magnetoconductance oscillations at high magnetic field.} Numerical derivative of the conductance as a function of the global back- and local bottom-gates at $B=$\SI{8}{T} where two additional sets fine oscillations can be observed (indicated with the green, dashed line)), superimposed on each set of cyan oscillations. Left and right side filling factors are indicated by green and purpler arrows, respectively.}
    \label{fig:Green_MCO}
\end{figure}

Two additional sets of magnetoconductance oscillations which that have been observed already in Fig.~\ref{fig:Bias_spectroscopy}c are shown in detail in Fig.~\ref{fig:Green_MCO}, marked in green. Detailed gate, magnetic-field, temperature and bias dependence is shown in the SI. The gate spacing is much shorter than for other osccilations (see SI). From magnetic field dependent measurements, spacing of $\Delta B=$\SI{6}{mT} at $B=$\SI{5.8}{T} to $\Delta B=$\SI{4}{mT} at $B=$\SI{8}{T} have been extracted. However, the magnetic field spacing of the second set of green magnetoconductance oscillations yields different values, ranging from $\Delta B=$\SI{25}{mT} at $B=$\SI{6}{T} to $\Delta B\sim$\SI{10}{mT} at $B=$\SI{8}{T}. The bias and temperature dependent measurements show vanishing oscillations around $V_\text{SD}\sim$\SI{\pm 1}{mV} and $T\sim$\SIrange{2}{3}{K}.

The origin of these oscillations is unknown. Aharonov-Bohm oscillations would correspond to an edge state spacing as high as \SI{\sim 700}{nm}, which is clearly unphysically large. The combination of edge states and a charge carrier island co-existing in the device could lead to Coulomb blockade oscillations. However, gate and magnetic-field spacings give inconsistent island sizes.  Details and more discussions are given in the SI. To resolve the origin of these oscillations further studies are needed.


\section{Conclusion}

In conclusion, we have observed three types of magnetoconductance oscillations in transport along a graphene \textit{p-n} junction.
We have demonstrated from the detailed analysis of the various oscillations that both snake states and Aharonov-Bohm oscillations appear within our measurements, and even co-exist in some parameter region. The question arises: how can a snake state, mostly imagined as a classical ballistic trajectory, exist in the regime where quantum effects also seem to be present, as demonstrated by Aharonov-Bohm interferences? Here we provide a comprehensive picture of both effects.

(1) By investigating the gate-gate maps we have seen that first, at the lowest densities and largest magnetic fields, Aharonov-Bohm oscillations originating from symmetry broken states appear. In this case no coupling between the edge states is present along the \textit{p-n} junction, only at the flake edges. At low doping the slope of the potential profile, and hence the electric field is small, which results in spatially separated edge channels which can only mix at the flake edge. This Aharonov-Bohm effect has been recently studied in Ref.~\cite{Wei17}, and modeled as two edge channels with different momentum along the \textit{p-n} junction. The Aharonov-Bohm flux can be calculated from the momentum difference of edge channels, as it is directly related to their guiding center \cite{Stegmann13}.

(2) As the bulk doping and hence the electric field is further increased, the edge-states propagating along the \textit{p-n} junctions are no longer eigenstates and start to mix. This effect has been studied previously for constant electric field, where it has been shown that for $E_c>v_F \cdot B$, mixing of the states occur, and electrons can cross the \textit{p-n} junction \cite{Lukose07, Shytov09, Shytov07}. In the very recent calculations of Cohnitz et al. in Ref.~\cite{Cohnitz16}, it has been shown, that in this regime interface modes, with velocity corresponding to classical snake states appear.
The real space motion of the center of the wavepackage, that gives rise to the snake movement along the \textit{p-n} junction, can thus be understood as an emergent spatial oscillations inherent in the modulus of the true energy eigenstate, which due to the mixing is a superposition of edge states on the left and right of the \textit{p-n} junction. In the simplest picture, one has a superposition of one mode on the left and one on the right that are coupled through the electric-field. This results into a periodic motion in real space mimicking snake-orbits with the effect of a periodic oscillations in the conductance which corresponds to the classical commensurability criterion. A simple model demonstrating this is given in the SI. The oscillation frequency depends on the potential strength and cyclotron frequency. The situation in our sample is more complex, since there are several channels and the electric field is position dependent: it is largest at the center of the \textit{p-n} junction and decreases further away from it. In addition, the magnetic field further complicates electrostatics due to the formation of Landau levels in the density of states. This makes quantitative analysis very challenging. Similar pictures based on numerical analysis have been presented in Refs. \cite{Kolasinski17,Milovanovic14}. Further details on this model will be given in the SI.

(3) Finally as the density is increased further other oscillations appear, marked by orange. We attribute them to Aharonov-Bohm oscillations between the lowest Landau levels, i.e. the inner-most edge states at the center of the \textit{p-n} junction. For these edge state on can find two interfering paths for which only one nearest-neighbor edge scattering along the graphene edge is needed to define an interference loop. Though higher order edge-states may contributes as well, there magnitude are much smaller, since to connect these states in an Aharonov-Bohm path that reaches from one side of the \textit{p-n} junction to other will require more than one scattering event giving rise to a very small visibility. Let us emphasize the different origins of the snake-state and Aharonov-Bohm oscillations: the former are caused by edge-state mixing in the bulk due the presence of a strong electric field, while the later rely on scattering along the graphene edge caused by edge disorder. We also stress that the orange magnetoconductance oscillations can be reproduced nearly perfectly using quantum transport simulations, without including electron-electron interactions or a Zeeman-term. Therefore we can exclude partial or full degeneracy lifting of the lowest Landau level in order to explain the orange oscillations.\\

Our study made the large steps in understanding the complex behaviour graphene \textit{p-n} junctions observed on the boundary of quasi-classical and quantum regime, and showed the surprising finding that Aharonov-Bohm like interferences and quasi-classical snake states can co-exist. Future studies might focus on the origin of the transition between tilted and checker-board pattern in the bias-dependence of the snake states.
In further steps interferometers based on bilayer graphene can be constructed, where electrostatic control of edge channels is possible. Recent works have shown the potential to engineer more complex device architectures \cite{Zimmermann2017, hiske_qpc}, where also exotic fractional quantum Hall states could be addressed. \\



\textbf{Acknowledgments}

The authors gratefully acknowledge fruitful discussions on the interpretation of the experimental data with Peter Rickhaus, Amir Yacobi, L\'aszl\'o Oroszl\'any, Reinhold Egger, Alina Mrenca-Kolasinska, Csaba T\H{o}ke and thank Andr\'as P\'alyi for discussion ons the quantum snake model presented in the SI.  This work has received funding from the European Union Horizon's 2020 research and innovation programme under grant agreement No 696656 (Graphene Flagship), the Swiss National Science Foundation, the Swiss Nanoscience Institute, the Swiss NCCR QSIT, ISpinText FlagERA network and from the OTKA PD-121052 and OTKA FK-123894 grants, and K.R and M.H.L. from the Deutsche Forschungsgemeinschaft (project Ri 681/13). P.M. acknowledges support from the Bolyai Fellowship. Growth of hexagonal boron nitride crystals was supported by the Elemental Strategy Initiative conducted by the MEXT, Japan  and  JSPS  KAKENHI  Grant  Numbers  JP26248061, JP15K21722,  and  JP25106006.



\end{document}


\beginsupplement

\title{Co-existence of classical snake states and Aharanov-Bohm oscillations along graphene \textit{p-n} junctions}


\author{P\'eter Makk}
\thanks{These authors contributed equally}
\email{Peter.makk@unibas.ch}
\affiliation{Department of Physics, University of Basel, Klingelbergstrasse 82, CH-4056 Basel, Switzerland}
\affiliation{Department of Physics, Budapest University of Technology and Economics and Condensed Matter Research Group of the Hungarian Academy of Sciences, Budafoki ut 8, 1111 Budapest, Hungary}

\author{Clevin Handschin}
\thanks{These authors contributed equally}
\affiliation{Department of Physics, University of Basel, Klingelbergstrasse 82, CH-4056 Basel, Switzerland}

\author{Endre Tovari}
\affiliation{Department of Physics, Budapest University of Technology and Economics and Condensed Matter Research Group of the Hungarian Academy of Sciences, Budafoki ut 8, 1111 Budapest, Hungary}

\author{Kenji Watanabe}
\affiliation{National Institute for Material Science, 1-1 Namiki, Tsukuba, 305-0044, Japan\\}

\author{Takashi Taniguchi}
\affiliation{National Institute for Material Science, 1-1 Namiki, Tsukuba, 305-0044, Japan\\}

\author{Klaus Richter}
\affiliation{Institut für Theoretische Physik, Universit\"{a}t Regensburg, D-93040 Regensburg, Germany}

\author{Ming-Hao Liu}
\affiliation{Department of Physics, National Cheng Kung University, Tainan 70101, Taiwan}

\author{Christian Sch\"{o}nenberger}
\email{Christian.Schoenenberger@unibas.ch}
\affiliation{Department of Physics, University of Basel, Klingelbergstrasse 82, CH-4056 Basel, Switzerland}

\date{\today}

\maketitle

\tableofcontents{}
\newpage

\section{Fabrication of pn-junctions with local bottom-gates}
\begin{figure}[htbp]
    \centering
      \includegraphics[width=1\textwidth]{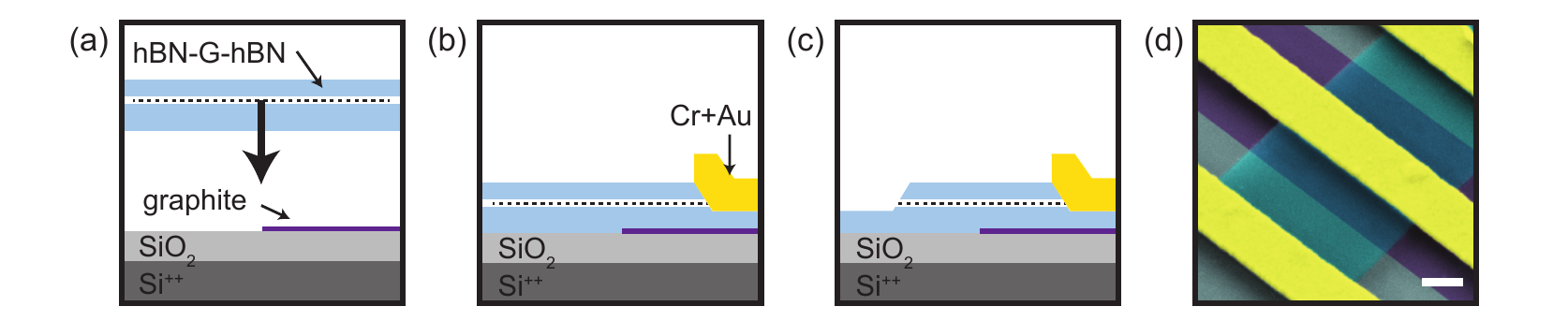}
    \caption{\textbf{Fabrication of a two-terminal \textit{p-n} junction array. a-c,} The assembly of the hBN-graphene-hBN heterostructure and establishing the side-contacts follows mostly the procedure described in \citep{Wang13}. \textbf{d,} False-color SEM image with leads indicated with yellow, the hBN-graphene-hBN heterostructure with cyan and the  bottom-gates with purple. Grey is the $SiO_2$-covered doped Si backgate. Scale-bar equals \SI{200}{nm}.}
    \label{fig:Fabricatioin}
\end{figure}

The results shown in the main text were measured on a graphene p-n junction encapsulated in hexagonal boron-nitride (hBN), as illustrated in Fig.~\ref{fig:Fabricatioin}a-d. The fabrication of the hBN-graphene-hBN heterostructure follows in most steps Ref. \citep{Wang13} with some variations and extensions as explained in the following. The graphene exfoliation is done on a  Si$^{++}$/SiO$_2$ substrate (SiO$_2$ is \SI{300}{nm} thick), using the scotch-tape technique. The chips were previously cleaned using Piranha solution (98\% H$_2$SO$_4$ and 30\% H$_2$O$_2$ in a ratio of 3:1). The full heterostructure is placed on a pre-patterned few-layer graphene (using a PMMA/HSQ mask and an O$_{2}$-plasma for etching) which is later used as a series of local bottom-gates, as shown in Fig.~\ref{fig:Fabricatioin}a. Next, self-aligned side-contacts are established to the graphene as shown in Fig.~\ref{fig:Fabricatioin}b. Self-aligned means that the same PMMA mask is used to etch down the hBN-graphene-hBN heterostructure and subsequently directly evaporate the Cr/Au (\SI{10}{nm}/\SI{50}{nm}) contacts. This leads to electrically very transparent contacts (\SI{\sim 400}{\ohm \micro m}) since the exposed graphene edge never comes into contact with any solvent or polymer. It is worth mentioning that we use cold-development ($T\sim$\SIrange{3}{5}{\degreeCelsius}) with IPA:H$_2$O (7:3) to reduce cracking of the PMMA on hBN \cite{Lee16_arxive,Rooks02}. The bottom hBN layer is not fully etched through (Fig.~\ref{fig:Fabricatioin}b), since otherwise it forms a short with the underneath lying bottom-gates. Finally, an etching step is required to define the device into \SI{1.5}{\micro m} wide channels, as illustrated in Fig.~\ref{fig:Fabricatioin}c. A false-color SEM picture is given in Fig.~\ref{fig:Fabricatioin}d.
For the fabrication details of the \textit{p-n-p} devices see Ref.~\cite{Handschin16}.

\newpage

\section{Characterization}
\begin{figure}[htbp]
    \centering
      \includegraphics[width=1\textwidth]{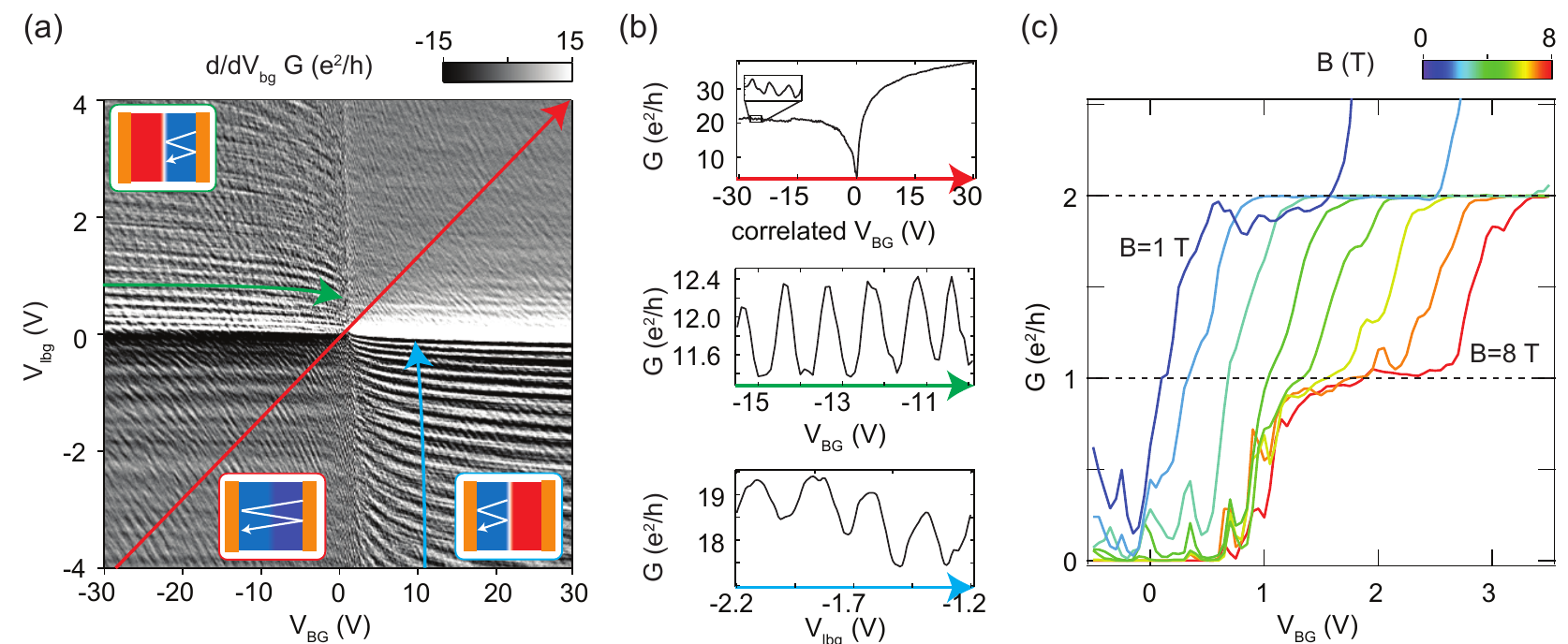}
    \caption{\textbf{Characterization of the two-terminal \textit{p-n} junction. a,} Numerical derivative of the conductance as a function of global back-gate ($V_{\text{BG}}$) and local bottom-gate ($V_{\text{lbg}}$) at zero magnetic field. Fabry-P\'{e}rot oscillations in between the leads and within the right and left cavity as illustrated in the inset are indicated with the red, blue and green arrows respectively. \textbf{b,} Representative linecuts within a (restricted) gate-range for of all three types of Fabry-P\'{e}rot oscillations as indicated by the colored arrows in (a). \textbf{c,} Quantum Hall measurement in the unipolar regime.}
    \label{fig:Characterization}
\end{figure}

Figure~\ref{fig:Characterization}a-b) shown Fabry-P\'{e}rot oscillations at zero magnetic field \cite{Campos12b,Grushina13,Rickhaus13,Varlet14,Shalom16,Calado15,Handschin16}, which are signatures of ballistic and phase-coherent transport between the leads. In Fig.~\ref{fig:Characterization}c quantum Hall measurements are shown. We observe partial degeneracy-lifting (formation of an insulating region $\nu=0$) for magnetic fields above \SIrange{2}{3}{T} and full degeneracy-lifting for magnetic-fields above \SIrange{5}{6}{T}.
It is worth noting that the potential-profile along the graphene edges (relevant for the quantum Hall effect) and the electrostatically defined \textit{p-n} junction (relevant for the Aharonov-Bohm interferometer) are quite different, because the potential profile at the graphene edge (hard wall) is relatively sharp compared to the much smoother potential profile of the \textit{p-n} junction. Therefore may be it is possible to observe signatures of full degeneracy lifting along the \textit{p-n} junction at magnetic fields being significantly lower as compared to full degeneracy lifting along the graphene edges.
Note that a fixed resistance (accounts for the contact- and line-resistance) was subtracted from the two-terminal measurement such that the conductance plateaus fit the expected values of $G=$2,6,10,...\si{e^2/h}.

\section{\textit{P-n} junctions in the proximity of the contacts}
\begin{figure}[htbp]
    \centering
      \includegraphics[width=1\columnwidth]{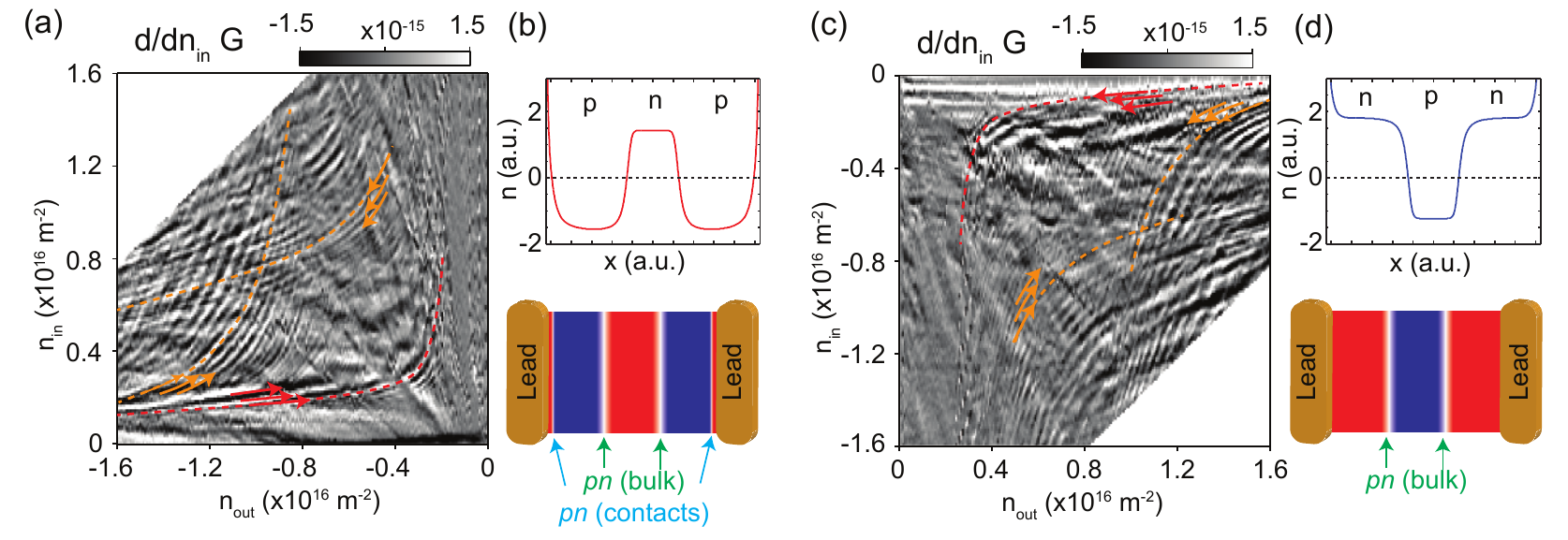}
    \caption{\textbf{Numerical derivative of the conductance in the bipolar regime of a \textit{p-n-p} device. a,} Configuration where the bulk is \textit{p-n-p} doped. The red and orange arrows indicate the two different types of magneto-conductance oscillations as shown in the main text. \textbf{b,} Illustration of the charge-carrier density profile along the device (top panel). The contacts dope the graphene in its proximity n-type, forming an additional \textit{p-n} junctions. The two types of \textit{p-n} junctions are indicated in the bottom panel. \textbf{c,} Same as in (a), but where the bulk graphene is \textit{n-p-n} doped. \textbf{d,} No additional \textit{p-n} junctions are formed near the contacts.}
    \label{fig:Proximitized_pn_Junction}
\end{figure}

In Fig.~\ref{fig:Proximitized_pn_Junction} we show that the two sets of magneto-conductance oscillations cannot be attributed to the two types of \textit{p-n} junctions which might be present in the system. The two possible types of \textit{p-n} junctions are: i) The \textit{p-n} junction formed in the bulk using the electrostatic gates as shown in Fig.~\ref{fig:Proximitized_pn_Junction}b,d. ii) The \textit{p-n} junction which is formed in the proximity of the contacts that electron-dope the graphene near them, if the bulk-graphene near the contacts is p-doped. In Fig.~\ref{fig:Proximitized_pn_Junction}a,c the two bipolar regimes of a gate defined \textit{p-n-p} device are shown. If the two sets of oscillations originated from the two types of \textit{p-n} junctions in the system, then two sets of oscillations would only be expected in the \textit{p-n-p} region, while in the n-p-n region one set would be absent. However, by comparing Fig.~\ref{fig:Proximitized_pn_Junction}a with Fig.~\ref{fig:Proximitized_pn_Junction}c it can be clearly seen that this is not the case. Note that features evolving nearly vertically vanish, since the numerical derivative was taken along the y-axis ($n_{\text{in}}$).
\newpage

\section{Skipping-length along a smooth \textit{p-n} junction}
\begin{figure}[htbp]
    \centering
      \includegraphics[width=1\columnwidth]{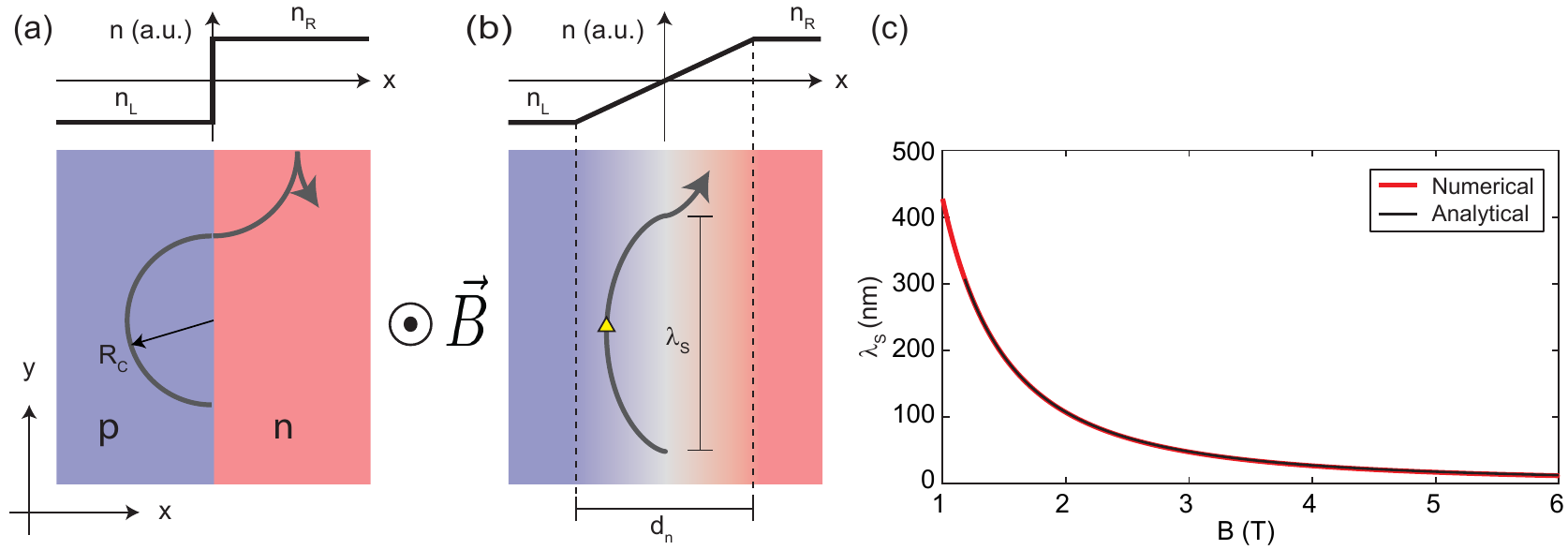}
    \caption{\textbf{Simplified and realistic snake state models. a,} Simplified snake state model including a step-function which models the \textit{p-n} junction. \textbf{b,} More realistic snake state model including a linear change within the distance $d_\text{n}$ between the charge carrier density $n_\text{L}$ and $n_\text{R}$ across the two sides of the \textit{p-n} junction. \textbf{c,} Comparison between analytically extracted skipping length $\lambda _\text{S}$ from equation~\ref{eq:lambda_s} in direct comparison with the values extracted from numerical calculations. Used parameters are $d_\text{n}=$\SI{200}{nm} and $n_\text{L}=n_\text{R}=$\SI{1e16}{m^{-2}}.}
    \label{fig:deriving_lambda_s}
\end{figure}

For the most simple snake state picture one considers a situation as sketched in Fig.~\ref{fig:deriving_lambda_s}a where the \textit{p-n} junction is modelled by a step-function where the absolute value of the density is constant, but opposite in sign on both sides of the \textit{p-n} junction. The cyclotron radius is therefore everywhere given by:
\begin{equation}\label{eq:Cyclotron}
R_\text{C}=\frac{\hbar k_\text{F}}{e B}
\end{equation}
where $k_\text{F}=\sqrt{|n|\pi}$.
However, a more realistic model as illustrated in  Fig.~\ref{fig:deriving_lambda_s}b includes a gradual change of the charge carrier density across the \textit{p-n} junction within the distance $d_\text{n}$. If the \textit{p-n} junction is centred at $x=0$, the bulk charge carrier density is given by $n_\text{L}$ and $n_\text{R}$ for $|x|>d_\text{n}/2$ while across the \textit{p-n} junction (for $|x|<d_\text{n}/2$) it is given by:
\begin{equation}\label{eq:slope}
n(x)=\frac{|n_\text{L}-n_\text{R}|}{d_\text{n}}x=S \cdot x
\end{equation}
where $S$ defines the slope. In the following we assume a positive slope as sketched in Fig.~\ref{fig:deriving_lambda_s}b and we concentrate on the regime where $|x|<d_\text{n}/2$.
The total force acting on a charge carrier is given by:
\begin{equation}\label{eq:Lorenz_force}
\vec{F}=\hbar \dot{\vec{k}}_\text{F}=q \left(\vec{E}+ \dot{r} \times \vec{B} \right),
\end{equation}
where $\dot{r}=(\dot{x},\dot{y},0)$, $\vec{E}=(E_\text{x},0,0)$ and $\vec{B}=(0,0,B_\text{z})$ as sketched in Fig.~\ref{fig:deriving_lambda_s}b. In graphene we can write $q=-\chi e$ ($e>0$ being the elementary charge and $\chi= \pm 1$ accounts for the electron ($+$) and hole ($-$) branch).
By explicitly working out $\dot{r} \times \vec{B}$, equation~\ref{eq:Lorenz_force} can be written component wise according to:
\begin{equation}\label{eq:F_xy}
\hbar \dot{k}_\text{x}=-\chi e \left(E_\text{x}+ \dot{y} B_\text{z} \right)  \qquad \text{and} \qquad  \hbar \dot{k}_\text{y}=\chi e \dot{x} B_\text{z}.
\end{equation}
Furthermore, using the energy dispersion relation of graphene ($E=\chi \hbar v_\text{F} k_\text{F}$), where $k_\text{F}=\sqrt{k_\text{x}^2+k_\text{y}^2}$, the values of $\dot{x}$ and $\dot{y}$ are given by:
\begin{equation}\label{eq:xy_dot}
\dot{x}=\frac{\partial E(k_\text{x},k_\text{y})}{\hbar\partial k_\text{x}}=v_\text{F} \frac{k_\text{x}}{k_\text{F}} \qquad \text{and} \qquad
\dot{y}=\frac{\partial E(k_\text{x},k_\text{y})}{\hbar\partial k_\text{y}}=v_\text{F} \frac{k_\text{y}}{k_\text{F}}.
\end{equation}
Based on the semiclassical equation of motions (equation~\ref{eq:F_xy} and equation~\ref{eq:xy_dot}) we now verify for a charge carrier starting perpendicular from the \textit{p-n} junction that:
\begin{itemize}
\item $k_\text{x}$ and $k_\text{y}$ perform a circular motion in k-space
\item the real-space trajectory of a charge carrier is given by a cycloid, if $|x|$ remains $<d_n/2$
\end{itemize}
To do so, the equation of motions are formed in such a way that the particles change their charge when crossing the junction, but not the group velocity.

\subsection{Circular motion of $k_\text{x}$ and $k_\text{y}$ in k-space}
To prove that $\dot{k}_\text{x}$ and $\dot{k}_\text{y}$ perform a circular motion in k-space we start by forming the ratio of the two according to $\dot{k}_\text{x}/\dot{k}_\text{y}=dk_\text{x}/dk_\text{y}$.
Using equation~\ref{eq:F_xy} where $\dot{k}_\text{x}$ and $\dot{k}_\text{y}$ are defined, as well equation~\ref{eq:xy_dot}, leads to:
\begin{equation}\label{eq:ratio}
\frac{\dot{k}_\text{x}}{\dot{k}_\text{y}}=-\frac{\left(E_\text{x} + v_\text{F} B_\text{z} \frac{k_\text{y}}{k_\text{F}} \right)}{v_\text{F} B_\text{z} \frac{k_\text{x}}{k_\text{F}}}.
\end{equation}
The electric field $E_\text{x}$ is directly related to the gradient of the band offset and thus the local Fermi-energy, and is given:
\begin{equation}\label{eq:E_x}
E_\text{x}=-\chi \frac{\hbar v_\text{F}}{e}\frac{dk_\text{F}}{dx}.
\end{equation}
where
\begin{equation}\label{eq:dk/dx}
\frac{dk_\text{F}}{dx}=\frac{\chi S \pi}{2 k_\text{F}}.
\end{equation}
having used $k_\text{F}=\sqrt{|n| \pi}=\sqrt{S |x| \pi}$.
Finally, plugging equation~\ref{eq:dk/dx} into equation~\ref{eq:E_x} leads to:
\begin{equation}\label{eq:E_x_final}
E_\text{x}=-\frac{\hbar v_\text{F}S \pi}{2 k_\text{F} e}.
\end{equation}
With equation~\ref{eq:E_x_final} we can now evaluate equation~\ref{eq:ratio}, leading to:
\begin{equation}\label{eq:ratio_2}
\frac{dk_\text{x}}{dk_\text{y}}=\frac{k_0-k_\text{y}}{k_\text{x}}
\end{equation}
where $k_0 \equiv (\hbar S \pi)/(2 e B_\text{z})$ is a constant. Solving equation~\ref{eq:ratio_2}
leads to:
\begin{equation}\label{eq:circular_k_space}
k_\text{x}^2+(k_\text{y}-k_0)^2=k_0^2
\end{equation}
which describes a circular motion in k-space. The integration constant is chosen in a way that the circle crosses the origin in k-space, corresponding to crossing the p-n junction ($x=0$, $k_F=0$) in real-space. From equation~\ref{eq:circular_k_space} we can learn that at the extremum of the trajectory (yellow triangle in Fig.~\ref{fig:deriving_lambda_s}b), $k_\text{x}=0$ and therefore $k_\text{y}=k_\text{F}=2 k_0$. We note that the local cyclotron radius (equation~\ref{eq:Cyclotron}) at the extremum is given by:
\begin{equation}\label{eq:Cyclotron_extremum}
R_\text{cyc}=\frac{\hbar 2k_0}{e B}=\left(\frac{\hbar}{e B}\right)^2 S \pi.
\end{equation}

\subsection{Cycloid motion in real space}
To verify that the circular motion in k-space describes indeed a cycloid motion in real space, we parametrize $k_\text{x}$ and $k_\text{y}$ on the hole side ($\chi=-1$) according to:
\begin{equation}\label{eq:kx_parameter}
k_\text{x}=k_0 \sin(\phi(t))
\end{equation}
and
\begin{equation}\label{eq:ky_parameter}
k_\text{y}=k_0 (1-\cos(\phi(t)))
\end{equation}
where the initial conditions are given by: $\phi(t=0)=0$, $k_\text{y}(t=0)=0$ which is increasing with time and $k_\text{x}(t=0)=0$ which is decreasing with time.
Assuming that the cycloid motion has the same parameter $\phi (t)$, the real-space trajectory is described by:
\begin{equation}\label{eq:x_parameter}
x=r (\cos(\phi(t)-1)
\end{equation}
\begin{equation}\label{eq:y_parameter}
y=r (\phi(t)-\sin(\phi(t)))
\end{equation}
and its derivative after $t$ is then given by:
\begin{equation}\label{eq:dotx_parameter}
\dot{x}=-r \dot{\phi}(t)\sin(\phi(t))
\end{equation}
\begin{equation}\label{eq:doty_parameter}
\dot{y}=r \dot{\phi}(t)(1-\cos(\phi(t)))
\end{equation}
The value of  $\dot{\phi}(t)$ is determined the following way: The value of $\hbar \dot{k}_\text{y}$ is given once by combining equation~\ref{eq:F_xy} and equation~\ref{eq:xy_dot}, and once by taking the time-derivative of equation~\ref{eq:ky_parameter} multiplied with $\hbar$. This leads to:
\begin{equation}
-e v_\text{F}\frac{k_\text{x}}{k_\text{F}}B_\text{z}=\hbar k_0 \sin(\phi(t))\dot{\phi}(t).
\end{equation}
By replacing $k_\text{x}$ with equation~\ref{eq:kx_parameter} this leads to:
\begin{equation}\label{eq:dot_phi}
\dot{\phi}(t)=\frac{e v_\text{F} B_\text{z}}{\hbar k_\text{F}}.
\end{equation}
Finally, by plugging
\begin{align*}
\sin(\phi(t))	\qquad &\rightarrow \qquad \text{from equation}~\ref{eq:kx_parameter}\\
(1-\cos(\phi(t)))	\qquad &\rightarrow \qquad \text{from equation}~\ref{eq:ky_parameter}\\
k_0 \qquad &\rightarrow \qquad \text{see below equation}~\ref{eq:ratio_2}\\
\dot{\phi}(t)  \qquad &\rightarrow \qquad \text{from equation}~\ref{eq:dot_phi}
\end{align*}
into the cycloid motion (equation~\ref{eq:dotx_parameter} and equation~\ref{eq:doty_parameter}) we find that the latter are equivalent with the equation of motion as given in equation~\ref{eq:xy_dot}, if:
\begin{equation}
r=\frac{R_\text{cyc}}{2},
\end{equation}
  where $R_\text{cyc}$is defined in equation~\ref{eq:Cyclotron_extremum}. This also means that at the extremum $|x| = R_{cyc}\sim S$ The skipping-length $\lambda _\text{S}$ in a cycloid is given by $\lambda _\text{S}=2 \pi r$, where $r$ is the radius of the ''rolling circle", therefore leading to:
\begin{equation}\label{eq:lambda_s}
\lambda _\text{S}=\left(\frac{\hbar \pi}{e B}\right)^2 S
\end{equation}
which is equivalent to the equation given in the main text. A direct comparison between numerical result (obtained by numerically solving the equations of motion) and the analytical formula equation~\ref{eq:lambda_s}) is shown in Fig.~\ref{fig:deriving_lambda_s}c, revealing an excellent agreement.

\subsection{Magnetic field spacing}
The magnetic field spacing can be extracted from equation~\ref{eq:lambda_s}. For simplicity we substitute $c\equiv (\pi \hbar)^2 S/e^2 $, such that it can be written as $\lambda _\text{S}=c/B^2$.
Undergoing one full oscillation period, one goes from $2 \lambda _\text{S} N = W$ to $2 \lambda _\text{S} (N+1) = W$. With equation~\ref{eq:lambda_s}, this can be rewritten as:
\begin{equation}
2 \frac{c}{B_\text{N}^2}N = W \qquad \text{and} \qquad 2 \frac{c}{B_\text{N+1}^2}(N+1) = W
\end{equation}
and consequently:
\begin{equation}
2 c = W (B_\text{N+1}^2-B_\text{N}^2)=W(B_\text{N+1}-B_\text{N})(B_\text{N+1}+B_\text{N}).
\end{equation}
For large $N$ one can approximate $(B_\text{N+1} - B_\text{N}) \sim \Delta B$ and $(B_\text{N+1} - B_\text{N}) \sim 2 B$. Finally, this leads to
\begin{equation}
\Delta B=\frac{c}{W}\frac{1}{B}
\end{equation}
which is equivalent to the equation in the main text.
\newpage

\subsection{Temperature dependence of snake states}

From equation~\ref{eq:lambda_s} it is possible to calculate the temperature dependence of snake states. At finite temperatures $T$ the Fermi-surface is broadened by $\Delta E\sim k_\text{B} T$ (where $k_\text{B}$ is the Boltzmann constant), thus leading to a spread of the Fermi-wavevector according to $\Delta k_\text{F}  \sim k_\text{B} T/(\hbar v_\text{F})$. The oscillations are expected to vanish if the smearing of trajectories becomes comparable to half a period:
\begin{equation}\label{eq:T_smearing_start}
2 \left( \lambda_\text{S,max}-\lambda_\text{S,min}\right) \cdot N \sim \left< \lambda_\text{S} \right>
\end{equation}
where $\lambda_\text{S,max}$ and $\lambda_\text{S,min}$ correspond to the maximal and minimal skipping-length due to the temperature smearing $\Delta k_\text{F}$, the average skipping-length $\left< \lambda_\text{S} \right>$ is given by equation~\ref{eq:lambda_s} and $N=W/(2\left< \lambda_\text{S} \right>)$. By using the relations $k_\text{F,max}^2-k_\text{F,min}^2 \sim 2k_\text{F}\Delta k_\text{F}$, equation~\ref{eq:T_smearing_start} can be rewritten as:
\begin{equation}\label{eq:T_smearing_Delta_k}
\Delta k_\text{F}\sim \frac{k_\text{F}^3}{W d_\text{n} \pi} \left(\frac{\pi \hbar}{e B} \right)^2.
\end{equation}
Together with $k_\text{F}=\sqrt{n \pi}$ one can rewrite equation~\ref{eq:T_smearing_Delta_k} to:
\begin{equation}\label{eq:T_smearing}
T_c \approx \frac{2 v_\text{F} \hbar^3}{W d_\text{n} k_\text{B} e^2 B^2} \sqrt{n^3 \pi ^5}.
\end{equation}

Here $T_c$ is the characteristic temperature, where the oscillations are expected to vanish.

\section{Temperature dependent measurements}
\begin{figure}[htbp]
    \centering
      \includegraphics[width=1\columnwidth]{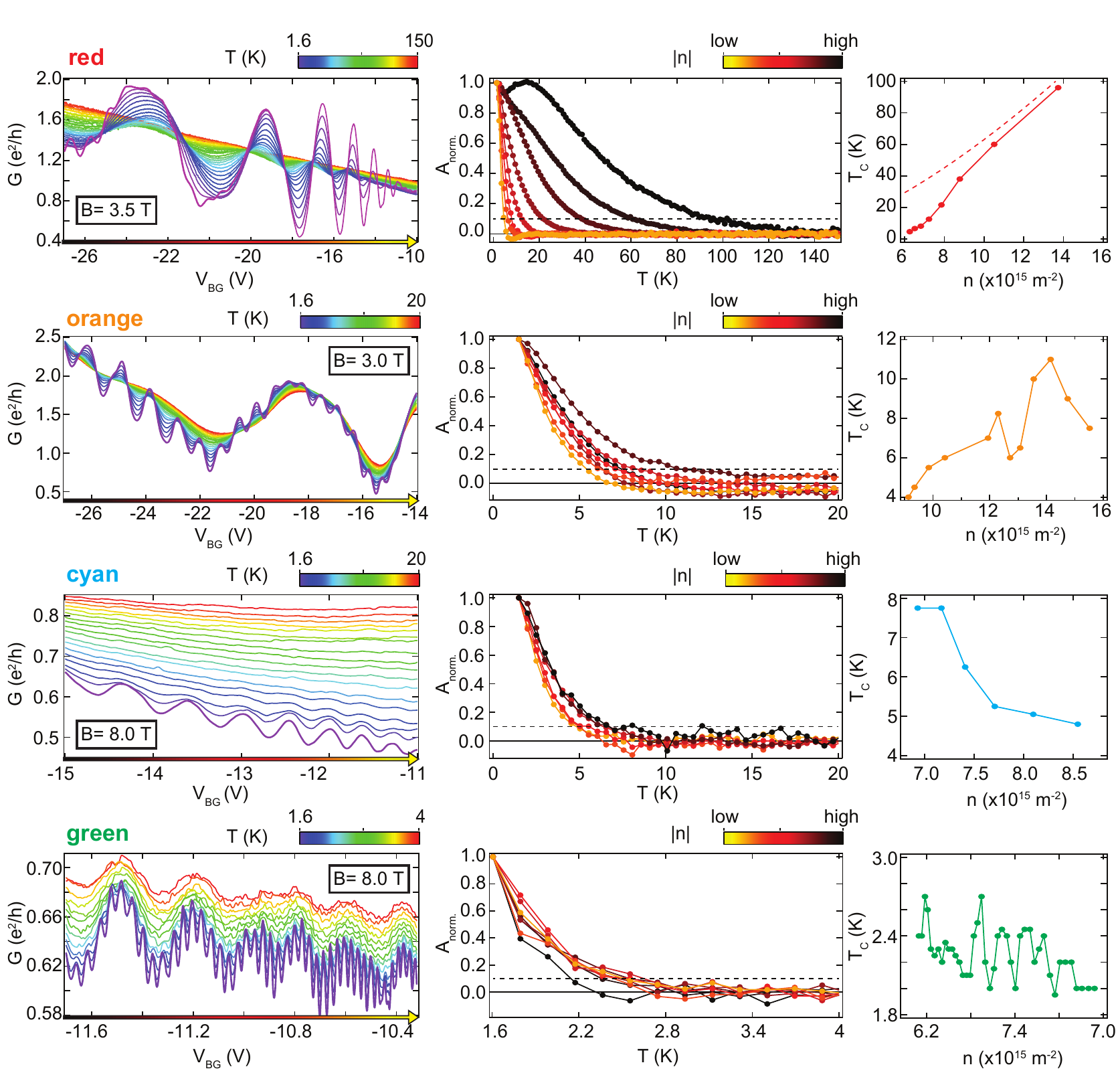}
    \caption{\textbf{Temperature dependence of the red, orange, cyan and green magnetoconductance oscillation. Left,} Conductance as a function of the global back-gates ($V_\text{lbg}$ is chosen such that $|n_\text{BG}| \sim |n_\text{lbg}|$) and temperature. \textbf{Middle,} Normalized area below a given oscillation as a function of temperature at a series of densities, the colors corresponding to the x axes of the plots of left column. \textbf{Right,} $T_{C}$ as a function of charge carrier density.}
    \label{fig:Temp_dep_extended}
\end{figure}
Fig.~\ref{fig:Temp_dep_extended} presents the data that has been used to extract the temperature dependence of the different types of magnetoconductance oscillations shown in the main text. In the left column, the conductance is given as a function of gate-voltage and temperature for a fixed magnetic field. The normalized area below a given oscillation:
\begin{equation}
A_\text{norm.}=\frac{A(T)}{A(T=1.6K)},
\end{equation}
is plotted in the middle column of Fig.~\ref{fig:Temp_dep_extended} as a function of temperature. The value $T_\text{C}$, which corresponds to the temperature at which the area under a given oscillation decreases to \SI{10}{\%} of its maximal value ($A_{\textrm{norm}}$), is then plotted on the right hand side of Fig.~\ref{fig:Temp_dep_extended} as a function of charge carrier density.

\section{Bias spectroscopy}
In this section details of the biasing models are given, which we used to calculate Fig.~5d-e in the main text, which is again shown here as Fig.~\ref{fig:Bias_spectroscopy} for convenience. We calculate the bias dependence for two different scenarios, namely the snake state model and an Aharonov-Bohm interference.\\
Upon applying a bias $V$ to the \textit{p-n} junction the total current $I$ is obtained by integrating $G(E)$, which is the energy-dependent conductance, over the bias window according to:
\begin{equation}\label{eq:integration}
I (V) \sim \int_{-aeV}^{(1-a)eV} G(E) dE
\end{equation}
where the parameter $a$ ($0<a<1$) defines the bias-window. Here we use $a$ to parameterize bias asymmetry between the two sides, which can be introduced as a shift of the bands with respect to the bias window. In the following we suppose that the bias drops predominantly over the \textit{p-n} junction. In this case the measured conductance is given by  $dI/dV$.

\subsection{Snake state interference}

We start with the oscillating conductance contribution of snakes states in a symmetric p-n junction:

\begin{equation}\label{eq:SS_conductance}
G(E) \sim \cos \left( \pi \frac{W}{\lambda_\text{S}} \right)= \cos \left( \pi \frac{W(eB)^2}{(\hbar\pi)^2 S} \right).
\end{equation}

Since most electrons are reflected off of the p-n junction, it is safe to assume that an applied DC bias drops over the p-n junction. It creates local equilibrium on either side, shifting the local chemical potential and thus the wave number. Using $\hbar k=\hbar k_\text{F} + E/v_\text{F}$ ($\hbar k_\text{F}$ is set by the electrostatic gates, $E/v_\text{F}$ is set by the applied bias and $E$ is measured with respect to $E_\text{F}$) and $S=(n_L-n_R)/d_n$, equation~\ref{eq:SS_conductance} can be rewritten as:

\begin{equation}\label{eq:SS_bias_2}
G(E) \sim \cos \left[ \frac{W e^2 d_\text{n} B^2}{\hbar^2 \left(k_\text{F}+\frac{a E}{2\hbar v_F}\right)^2+\hbar^2 \left(k_\text{F}+\frac{(1-a)E}{2\hbar v_F}\right)^2}\right].
\end{equation}

\begin{figure*}[tb]
    \centering
      \includegraphics[width=\columnwidth]{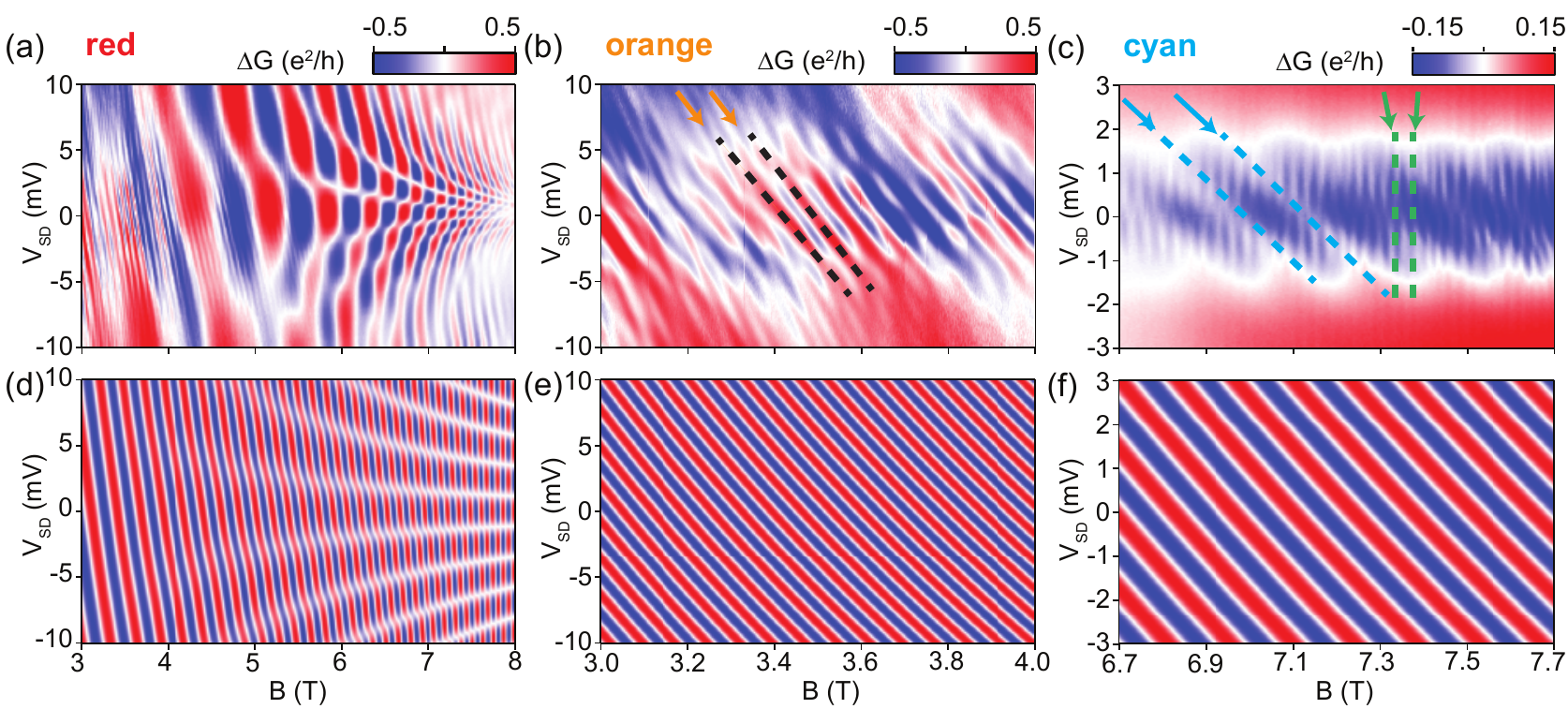}
    \caption{\textbf{Bias spectroscopy. a-c} Measurement of the red, orange and cyan magnetoconductance oscillations as a function of bias and magnetic field where a smooth background was subtracted. \textbf{d-f,} Simulations are corresponding to the measurements (a-c).  Parameters used: $W=$\SI{1.5}{\micro m}, $d_\text{n}=$\SI{100}{nm} (red oscillations), $k_\text{F}$ corresponding to $n\sim$\SI{1.7e12}{cm^{-2}} (red, orange) or $n\sim$\SI{0.8e12}{cm^{-2}} (cyan). For the Aharonov-Bohm oscillations we considered a bias dependent gating effect with $\alpha=$\SI{0.32}{nm/mV_\text{SD}} and $d=$\SI{40}{nm} (orange oscillation) or $\alpha=$\SI{0.25}{nm/mV_\text{SD}} $d=$\SI{20}{nm} (cyan oscillations), while a renormalization of the edge state velocity is neglected ($\beta=1$).}
    \label{fig:Bias_spectroscopy}
\end{figure*}

Assuming $E \ll \hbar k_\text{F} v_\text{F}$, which is reasonable for the applied bias, this leads to:
\begin{equation}\label{eq:SS_bias_3}
G(E) \sim \cos \left[ \frac{W e^2 d_\text{n} B^2}{2 (\hbar k_\text{F})^2} - \frac{W e^2 d_\text{n} B^2}{(\hbar k_\text{F})^3 v_\text{F}} E \right].
\end{equation}
Next, we define the parameters:
\begin{equation}\label{eq:subsitution_SS}
b \equiv \frac{W e^2 d_\text{n}}{2 (\hbar k_\text{F})^2} B^2  \qquad \text{and} \qquad  c \equiv \frac{W e^2 d_\text{n}}{(\hbar k_\text{F})^3 v_\text{F}} B^2.
\end{equation}
Upon applying a finite bias (equation~\ref{eq:integration}), the measured conductance is then obtained as:
\begin{align}\label{eq:SS_bias_general}
\frac{dI}{dV}\left(V\right) \sim &\Big\{ (1-a) \cos \big [-(1-a) \cdot e V \cdot c + b  \big ] \nonumber  \\
&+ a \cdot \cos \big [a \cdot e V \cdot c + b \big ] \Big\}.
\end{align}
In the case of a fully asymmetric biasing ($a=0$ or $a=1$), equation~\ref{eq:SS_bias_general} simplifies to:
\begin{equation}\label{eq:SS_bias_asymmetric}
\frac{dI}{dV}\left(V\right)\sim  \cos \big ( e V \cdot c + b  \big )
\end{equation}
reproducing the tilted pattern which is shown in Fig.~\ref{fig:Bias_spectroscopy}d at low magnetic field. On the other hand, for the case of completely symmetric biasing ($a=0.5$) equation~\ref{eq:SS_bias_general} simplifies to:
\begin{equation}\label{eq:SS_bias_symmetric}
\frac{dI}{dV}\left(V\right) \sim  \cos \big ( b \big ) \cdot \cos \big ( e V  \cdot c \big )
\end{equation}
which leads to the checker-board pattern which is shown in Fig.~\ref{fig:Bias_spectroscopy}c at high magnetic field. The checker-board pattern is in agreement with previous studies \cite{Morikawa15, Wei17} where a similar behaviour was observed. The oscillation period ($\Delta B$) decreases with increasing magnetic field in the simulation (Fig.~\ref{fig:Bias_spectroscopy}d) similarly to the experiment (Fig.~\ref{fig:Bias_spectroscopy}a). In order to reproduce the transition from tilted (asymmetric biasing, equation~\ref{eq:SS_bias_asymmetric}) to checker-board pattern (symmetric biasing, equation~\ref{eq:SS_bias_symmetric}), the parameter $a$  was varied linearly from $1 \rightarrow 0.5$ by going from low to high magnetic field. While the increasing magnetic field seems to be responsible for the transition from $a=1$ to $a=0.5$, the precise reason remains unknown so far. We speculate that it might be related to the capacitances in the system, such as the capacitance related to the insulating region with $\nu=0$ or the quantum capacitance from the bulk. The latter is directly proportional to the DOS \cite{Xia09,Yu13} and thus changes significantly upon tuning the Fermi energy from a Landau level into a Landau level gap. In contrast to the simulation, the checker-board pattern in the experiment vanishes upon increasing the bias. This has been attributed to a dephasing rate of the charge carriers being proportional to the bias-voltage \cite{Chamon97,Ji03,Wiel03,Roulleau08,McClure09}. However, this does not seem to apply for the tilted pattern, which persists up to $V_\text{SD}=$\SI{\pm 10}{mV}.

\subsection{Aharonov-Bohm interference}
Next we calculate the bias dependence for an Aharonov-Bohm interferometer. The magnetoconductance oscillations of a slightly modified model are given by:
\begin{equation}\label{eq:AB_bias_1}
G(E) \sim \cos \left[2 \pi\frac{W \cdot (d + \alpha E/e) \cdot B}{\Phi_0} + k(E) \Delta L \right].
\end{equation}
Here the parameter $\alpha$ is a phenomenological parameter in order to account for a bias dependent gating effect \cite{Bieri09}, which we will elaborate on further below. For simplicity the edge state spacing $d$ changes linearly with the applied bias. The factor $k \Delta L$ in equation~\ref{eq:AB_bias_1} accounts for a possible path-difference between the edge states, where $k$ is replaced by $k=k_\text{F}+ E/(\hbar v_\text{F} \beta)$. The parameter  $\beta$ ($0\leq \beta \leq 1$) was introduced to account for the renormalized edge state velocity compared to the Fermi velocity \cite{Cohnitz16}.
Solving the bias-dependence for equation~\ref{eq:AB_bias_1} leads again to equation~\ref{eq:SS_bias_general}, however the coefficients $b$ and $c$ are now given by:
\begin{equation}\label{eq:subsitution_AB}
b \equiv  \frac{2 \pi W d}{\Phi _0} B + k_\text{F} \Delta L \qquad \text{and} \qquad c \equiv   -\frac{\Delta L}{\hbar v_\text{F} \beta}-\frac{2 \pi W  \alpha}{\Phi _0}B.
\end{equation}
We have used asymmetric biasing, $a=1$ for the simulations, since no transition to checkerboard pattern was observed for Aharonov-Bohm oscillations.
We have found that the second term alone ($\alpha=0$) of eq.~\ref{eq:AB_bias_1} cannot lead to substantial bias dependence if the parameters are chosen realistically ($\Delta L\sim$\SI{20}{nm} and $\beta =1$). To account for the tilt of the measurement a considerable renormalization of the edge state velocity would be needed, leading to an unphysically large reduction of $v_F$ by a factor of one hundred. Therefore most of the tilt must come from non-zero $\alpha$: the bias induced gating effect.

\subsubsection{Bias dependent gating effect for Aharanov-Bohm interferences}
\begin{figure}[htbp]
    \centering
      \includegraphics[width=1\columnwidth]{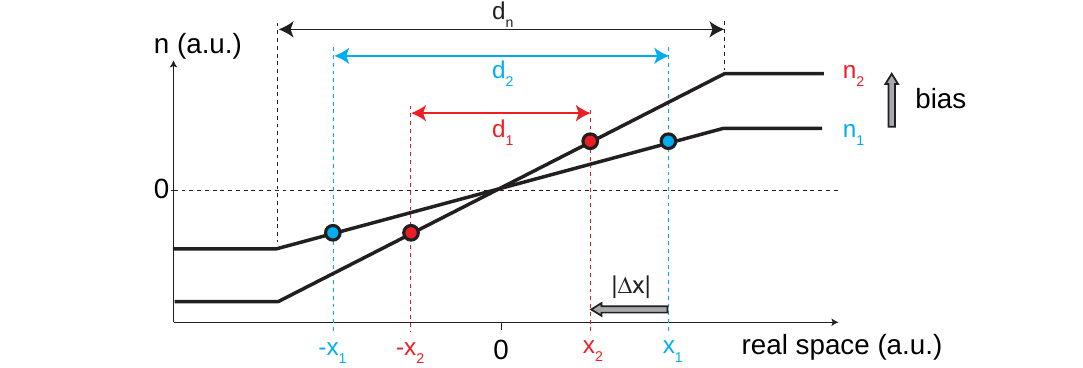}
    \caption{\textbf{Bias dependent gating effect. } By applying a finite bias, the charge charier density in the bulk changes, consequently leading to a different slope across the \textit{p-n} junction (if $d_\text{n}$ remains fixed). This then leads to a shift of the real space location of the edge state (red and blue dots).}
    \label{fig:AB_bias_model}
\end{figure}

In the following we assume that the Aharonov-Bohm interferometer is located in the center of the \textit{p-n} junction around the insulating region ($\nu=0$), and we consider a symmetric \textit{p-n} junction for simplicity. The slope $S_1$ is given by $S_1=2n_1/d_\text{n}$, and $n_1$ corresponds to the charge carrier doping induced by the electrostatic gates.
By symmetrically applying a bias of energy $E=\pm e V_\text{SD}/2$ at each side of the \textit{p-n} junction the local chemical potential changes the same amount, and the bulk density increases from $n_1$ to:
\begin{equation}\label{eq:AB_bias_estimate}
n_2=\frac{\left(\sqrt{|n_1| \pi}+\frac{eV_\text{SD}}{2 \hbar v_\text{F}}\right)^2}{\pi}
\end{equation}
as illustrated in Fig.~\ref{fig:AB_bias_model} with the grey arrows. Here we assumed a perfect transmission from contacts into graphene, neglecting the capacitive coupling between them. The increase of the bulk density leads to a shift of the real space position of the edge states by $\Delta x$ as indicated with the red and blue dots. Assuming that at $n_1$ the edge states are located at $\pm x_1$ (symmetric \textit{p-n} junction), then the new locations are given by:
\begin{equation}
\pm x_2=\pm x_1 \frac{S_1}{S_2}.
\end{equation}
Finally, the shift of the edge state is given by $|\Delta x|=|x_1-x_2|$, or:
\begin{equation}\label{eq:AB_bias_estimate_final}
|\Delta x|=x_1 \left| 1-\frac{n_1}{n_2} \right|,
\end{equation}
where $S \propto n$ ($d_\text{n}=\text{const.}$) was used. Assuming $eV \ll \hbar v_F \sqrt{|n|\pi}$, then  $|\Delta x| \propto V$ as proposed in Eq.~\ref{eq:AB_bias_1}. Here as a further simplification we have neglected the magnetic field dependence of $\alpha$. We note here that the same result can be obtained using arbitrary biasing asymmetry factor, $a$. \\
For the orange magnetoconductance oscillations the gate induced bulk density was given by $n_1 \sim$\SI{1.7e16}{m^{-2}}, and the bias spacing was extracted to $V_\text{SD}\sim$\SI{3}{mV} from the experiment. Using equation~\ref{eq:AB_bias_estimate}, then $n_2$ is given by $n_2 \sim$\SI{1.75e16}{m^{-2}} (at $V_\text{SD}\sim$\SI{3}{mV}). Based on the magnetic field spacing of the magneto-oscillations (not shown) we estimate $x_1\sim$\SIrange{23}{28}{nm}. Note that the change of the edge state spacing is given by $|2\Delta x|$, leading to $|2\Delta x|\sim$\SIrange{1.19}{1.44}{nm}, or $\alpha \sim$\SIrange{0.39}{0.48}{nm/mV_\text{SD}}.  The measurements yield a comparable value, $\alpha \sim$\SI{0.32}{nm/mV_\text{SD}}.

For the cyan magnetoconductance oscillations the gate induced bulk density was given by $n_1 \sim$\SI{9e15}{m^{-2}}, and the bias spacing was extracted to $V_\text{SD}\sim$\SI{1.5}{mV} from the experiment. Then $n_2 \sim$\SI{9.12e15}{m^{-2}} (at $V_\text{SD}\sim$\SI{1.5}{mV}) and we estimate $x_1\sim$\SIrange{9}{14}{nm}. This leads to $|2\Delta x|\sim$\SIrange{0.23}{0.37}{nm}, or $\alpha \sim$\SIrange{0.16}{0.25}{nm/mV_\text{SD}}. The measurements yield $\alpha \sim$\SI{0.25}{nm/mV_\text{SD}}. These estimates agree fairly well with the measurement considering the simple nature of our estimate.

\section{Aharonov-Bohm interferences based on electrostatic simulations}

\begin{figure}[tb]
    \centering
      \includegraphics[width=1\textwidth]{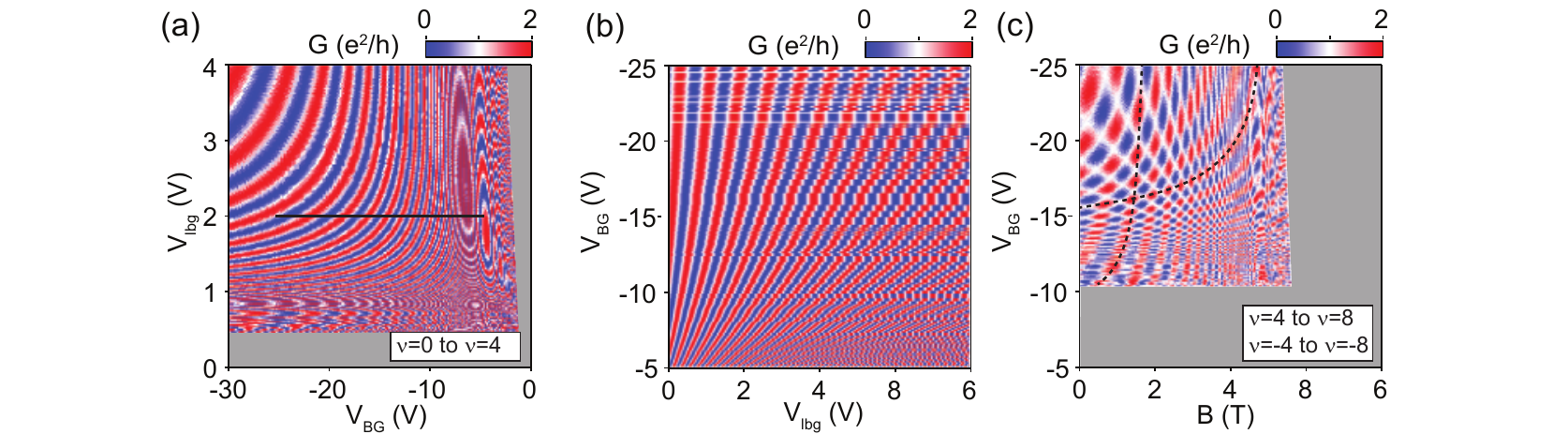}
    \caption{\textbf{Edge state interference between neighbouring Landau levels. a,} Conductance as a function of the global back-gate and local bottom-gate according to equation~\ref{eq:ES_interference} for the interference between the edge states with $\nu=0$ and $\nu=4$. At low doping (shaded in gray) $\nu=4$ is not populated. \textbf{b,} Conductance as a function of magnetic field for the linecut as indicated in (a) with the solid, black line. \textbf{c,} Interference between the edge states with $\nu=4$ and $\nu=8$ superimposed on top of the interference between edge states with $\nu=-4$ and $\nu=-8$ at $B=$\SI{2}{T}. At low doping (shaded in gray) one or several of the edge states are not populated.}
    \label{fig:Electrostatic_Simulation}
\end{figure}

The knowledge of the edge state position and their relative spacing ($d$) would allow to calculate the Aharanov-Bohm interference. We can formulate a phenomenological model according to:
\begin{equation}\label{eq:ES_interference}
G=G_\text{bgr} + G_\text{osc} \cdot \cos \left( 2\pi \frac{d \cdot W \cdot B}{\Phi _0}  \right),
\end{equation}
where $G_\text{bgr}$ and $G_\text{osc}$ define the background and oscillation amplitude, $W$ is the width of the graphene device (length of the \textit{p-n} junction) and the cosine accounts for a smooth oscillation. To extract the spacing $d$, the real-space position of the edge states moving along the \textit{p-n} junction was calculated from the density profile perpendicular to the \textit{p-n} junction. For simplicity we used the density profile $n(x)$ which was calculated for the case of zero magnetic field. In the simulation, for every set of ($V_{\text{BG}}$,$V_{\text{TG}}$) a density profile parallel with the x-axis (defined perpendicular to the \textit{p-n} junction, with $x=0$ centred in the middle of the top-gate) was calculated based on the quantum capacitance model for graphene \cite{Liu13_2} with classical self-partial capacitances simulated using FEniCS \cite{FEniCS} and Gmsh \cite{Gmsh}.  The real-space position of the edge states was extracted using the definition of the filling factors which defines the ratio between charge carrier density and magnetic field according to $n=\nu e B /h$. The conductance resulting from the interferences between the area enclosed by the edge states with $\nu=0$ and $\nu=4$ is shown in Fig.~\ref{fig:Electrostatic_Simulation}a as a function of the global back-gate and local bottom-gate at $B=$\SI{2}{T}. Furthermore, the magnetic field dependence of a selected linecut is shown in Fig.~\ref{fig:Electrostatic_Simulation}b. The magnetoconductance oscillations follow a roughly parabolic trend with increasing magnetic field strongly resembling the experimental results shown in the main text.
Upon superimposing the interference between two sets of edge states (e.g. $\nu=0$, $\nu=4$ and $\nu=0$, $\nu=-4$) one can qualitatively reproduce the two sets of orange magnetoconductance oscillations which are shifted in doping (not shown). This is in good agreement with the experimental results and the results from the quantum transport calculations which are both shown in the main text.


\section{Charging effects}
\begin{figure*}[h!]
    \centering
      \includegraphics[width=1\textwidth]{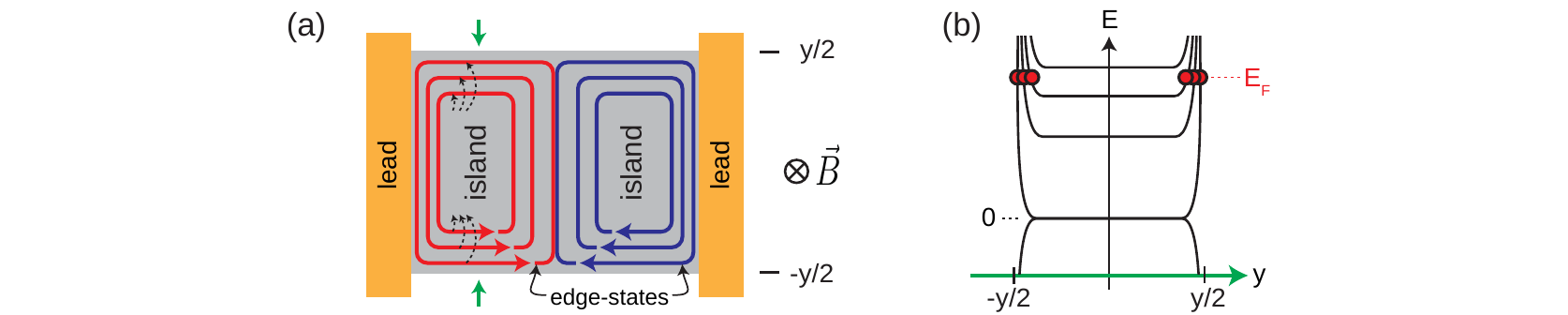}
    \caption{\textbf{Charging effects in a graphene \textit{p-n} junction at finite magnetic field. a,} Formation of a charge carrier island  which is capacitively coupled to the edge states (red/blue lines). The transfer of charge carriers between the edge states and the island is indicated for the n-doped island with the black, dashed arrows. \textbf{b,} Energy dispersion along a linecut as indicated in (a) with the green arrows. A combination between edge states (red dots) and a conducting island in the center of the device is present.}
    \label{fig:Charging_Effects}
\end{figure*}

Here the influence of charging effects shall be briefly discussed for the situation sketched in Fig.~\ref{fig:Charging_Effects}a.
In the bipolar regime and upon applying a magnetic field, the charge carriers move in edge states along the graphene edge and the \textit{p-n} junction. They can propagate along contact edges as well, if transmission to the contacts is low, and edge states are mostly reflected. Thus chargeable islands of partially filled Landau levels can form in the bulk of the device, tunnel-coupled to the edge states that encircle them.
As derived in Ref.~\cite{Rosenow07} the magnetoconductance oscillation for the charging model is given by:
\begin{equation}\label{eq:charging_conductance}
G \sim \cos \left[- 2 \pi \frac{\si{\Phi}}{\si{\Phi}_0} + 2 \pi \frac{\Delta _\text{X}}{\Delta} \left ( \tilde{\nu}\frac{\si{\Phi}}{\si{\Phi}_0} + N - N_\text{gate}  \right ) \right ],
\end{equation}
where $\si{\Phi}$ and $\si{\Phi}_0$ correspond to the magnetic flux through the area $A$ and the magnetic flux quantum respectively, $\Delta _\text{X}$ describes the coupling energy for one extra electron on the island, $\Delta$ is the level spacing, $\tilde{\nu}$ is the number of completely filled edge states, and $N$ and $N_\text{gate}$ give the number of charges on the island, and the number attracted by the gates, respectively. While the values of $\tilde{\nu}$ and $\Delta=2 \pi \hbar v_\text{F}/L$ \cite{Rosenow07}, where $L$ is the total interference path surrounding the island, are known, we do not have any knowledge of $\Delta _\text{X}$.

However, from Quantum Hall effect measurements, we know that our edge states are strongly coupled to contact, therefore this scenario is likely not occuring in our sample.

\section{Green oscillations}

\begin{figure*}[tb]
    \centering
      \includegraphics[width=1\columnwidth]{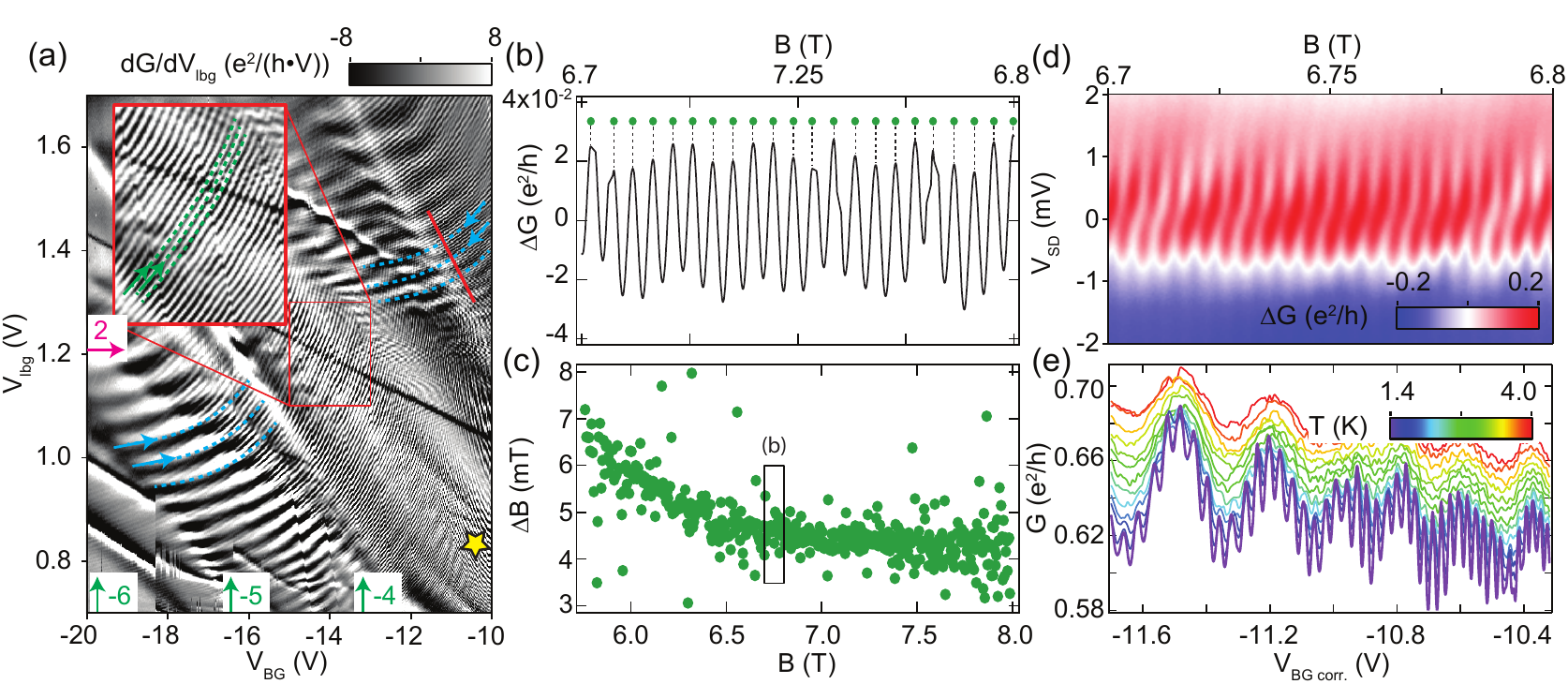}
    \caption{\textbf{Additional magnetoconductance oscillation at high magnetic fields. a,} Numerical derivative of the conductance as a function of the global back- and local bottom-gate at $B=$\SI{8}{T} where two additional fine oscillations can be observed (indicated for one set with the green, dashed lines). \textbf{b,} Background-subtracted oscillation of the conductance $\Delta G$ as a function of magnetic field within a limited field-range. The peak-positions are indicated with the green dots. \textbf{c,} Magnetic field spacing ($\Delta B$) extracted from a cut, which included the one in panel (b). \textbf{d,} Green magnetoconductance oscillations as a function of bias and magnetic field where a smooth background was subtracted. In panels (b-d) the gate-voltage remained fixed (indicated with the yellow  star in (a)). \textbf{e,} Temperature dependence of the green magnetoconductance oscillations which vanish around \SIrange{2}{3}{K}.}
    \label{fig:Green_MCO}
\end{figure*}

We now briefly describe and discuss the characteristics of the additional set of magnetoconductance oscillations labelled with green.
In Fig.~\ref{fig:Green_MCO}a the gate-gate dependence of the green magnetoconductance oscillations is shown in the bipolar regime, revealing a gate spacing within the measured gate range from $V_\text{BG}\sim$\SI{40}{mV} at lower doping up to $V_\text{BG}\sim$\SI{150}{mV} at higher doping. In Fig.~\ref{fig:Green_MCO}b the magnetic field dependence at a fixed gate-gate configuration, as indicated in Fig.~\ref{fig:Green_MCO}a with the yellow star, is shown. From the latter a magnetic field spacing of $\Delta B=$\SI{6}{mT} at $B=$\SI{5.8}{T} to $\Delta B=$\SI{4}{mT} at $B=$\SI{8}{T} was extracted as shown in Fig.~\ref{fig:Green_MCO}c. However, the magnetic field spacing of the second set of green magnetoconductance oscillations, which is not shown here, yields different values ranging from $\Delta B=$\SI{25}{mT} at $B=$\SI{6}{T} to $\Delta B\sim$\SI{10}{mT} at $B=$\SI{8}{T}.
The bias and temperature dependencies are furthermore shown in Fig.~\ref{fig:Green_MCO}d,e where a vanishing of the green magnetoconductance oscillations is seen around $V_\text{SD}\sim$\SI{\pm 1}{mV} and $T\sim$\SIrange{2}{3}{K}.\\
From the narrow gate spacing we first suspected that the green magnetoconductance oscillations originate from a charging effect (see previous section) \cite{Rosenow07}. In the latter edge states and charge carrier islands co-exist in the device. Such systems have been investigated in 2DEG Fabry-P\'{e}rot interferometers \cite{Zhang09_3,McClure09}. Note that the formation of a charge carrier island requires edge channels which are weakly coupled to the leads. However, our system is more likely in the strong coupling regime. This assumption is based on the observation of integer quantum Hall plateaus which are well developed throughout the whole unipolar regime. One could imagine disorder defined quantum dots, however, the good quality of the sample and the large area of these dots ($10^4-10^5\,$nm$^2$, based on their gate voltage spacing and the known capacitances) makes this scenario unphysical.

In addition, the area extracted from the gate spacing does not match the area extracted from the magnetic field spacing (simple Aharonov-Bohm flux estimation). The green magnetoconductance oscillations are furthermore not parallel with respect to the charge neutrality lines of the left and right sides of the \textit{p-n} junction. This we would naively expect for a charging effect where each cavity is predominantly tuned by one of the two gates. A possible explanation might be the shift of the \textit{p-n} junction position \cite{Handschin16}, which depends on both $V_\text{BG}$ and $V_\text{lbg}$, consequently changing the area of the two cavities.

Even though the charging model does not fit the experiments in various points, an Aharonov-Bohm oscillation also cannot explain the green magnetoconductance oscillations either. This is because the resulting edge state spacing extracted from the two sets would be very different (and unreasonably large), namely \SI{\sim 700}{nm} and \SI{\sim 150}{nm}. This contradicts the device geometry where the two cavities have roughly the same dimensions. To resolve these inconsistencies further studies will be needed.

\section{Snake states as coherent oscillations - Parallel wire, a minimal model}

\begin{figure}[tb]
    \centering
      \includegraphics[width=1\textwidth]{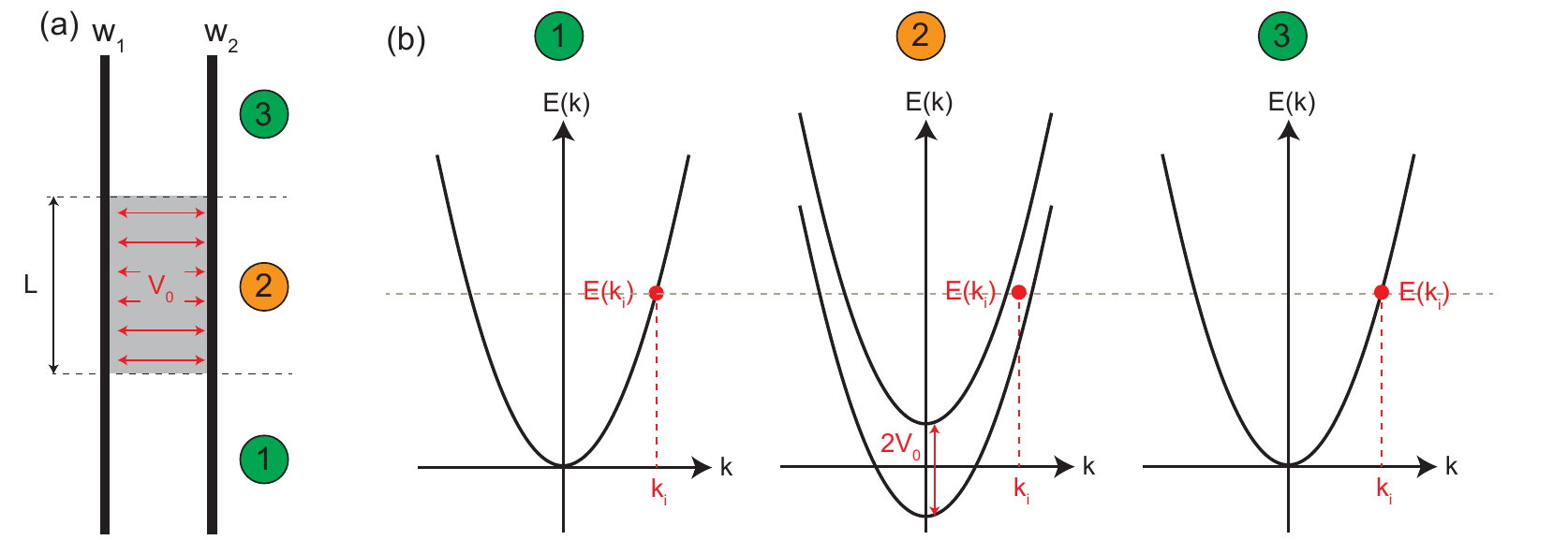}
    \caption{\textbf{Minimal model based on two parallel wires. a,} The two wires $w_1$ and $w_2$ are only coupled in region $2$ via a potential $V_0$, while otherwise they are decoupled from each other (in region $1$ and $2$. \textbf{b,} Dispersion relations of the wires in region $1-3$. In region $1$ and $2$ the dispersion relation is degenerate for the two wires.}
    \label{fig:Coherent_oscillation}
\end{figure}

Here we present a minimal model for the snake states. In Fig.~\ref{fig:Coherent_oscillation} a sketch of the model is shown: it consists of two 1-dimensional wires ($w_1$ and $w_2$) with free charge carriers, as labelled in Fig.~\ref{fig:Coherent_oscillation}a. While in regions $1$ and $3$ the two wires are decoupled from each other, they are coupled in region $3$ (having the length $L$) via the potential $V_0$.
We choose the following basis:
\begin{equation}
w_1=\begin{pmatrix}
  1\\
  0\\
\end{pmatrix}
\qquad
\text{and}
\qquad
w_2=\begin{pmatrix}
  0\\
  1\\
\end{pmatrix},
\end{equation}
where the wave-function describing a charge carrier incoming on $w_1$ is given by:
\begin{equation}
\Psi_1=e^{ikx}\begin{pmatrix}
  1\\
  0\\
\end{pmatrix}.
\end{equation}
In region $2$, the eigenstates are given by:
\begin{equation}
w_1 \pm w_2=e^{ikx}\frac{1}{\sqrt{2}}\begin{pmatrix}
  1\\
  \pm 1\\
\end{pmatrix},
\end{equation}
while in region $3$ the eigenstates are equivalent to region $1$ again.
In region $1$ and $3$ the dispersion relation corresponds to a degenerate free-electron dispersion:
\begin{equation}
E(k)=\frac{\hbar^2 k^2}{2m},
\end{equation}
whereas in region $2$, where the two wires are hybridized, it is split by the potential $V_0$:
\begin{equation}
E_{\pm}(k)=\frac{\hbar^2 k^2}{2m}\pm V_0.
\end{equation}

An electron injected from region $1$, with momentum $k_\text{i}$ and energy $\frac{\hbar^2 k_\text{i}^2}{2m}$ is not in an egienstate in region 2, therefore it will follow a coherent precession between the two wires resulting in an oscillation between the eigenstates $(1,0)$ and $(0,1)$. Depending on the length of the coupling region $2$ ($L$) and on the oscillation frequency $\omega=2V_0/\hbar$, the state ends up in either $w_1$ or $w_2$, that is in wire 1 or 2. This corresponds to the picture of the classical snake states. We expect an oscillation in the conductance as a function of both gate voltages and the magnetic field, as they affect the electric field across the p-n junction and also the spacing of edge states (which are no longer eigenstates), and therefore the coupling $V_0$. If the states in region $1$ and $3$ are split by e.g. an external magnetic field ($\hbar \omega_c$), and thus belong to different Landau levels, than the oscillation frequency will depend both on the coupling and the original splitting.

This model is analogous of a spin 1/2 model, where the spin in the beginning is prepared in the z-direction, and than a magnetic field in the x-direction is turned on. This will start to precess the spin around the x-axis with the Larmor frequency, and the measurement in the z-direction will show an oscillatory behaviour as a function of time (as naively already expected from spin-precession).